\documentclass[journal, comsoc, final]{IEEEtran}
\usepackage{graphicx}
\usepackage{booktabs}
\usepackage{amsmath}
\usepackage{array}
\usepackage{tabularx}
\newcolumntype{P}[1]{>{\centering\arraybackslash}p{#1}}
\newcolumntype{M}[1]{>{\centering\arraybackslash}m{#1}}
\newcolumntype{L}[1]{>{\centering\arraybackslash}l{#1}}
\newcolumntype{Z}{>{\centering\let\newline\\\arraybackslash\hspace{0pt}}X}

\newcolumntype{Y}[1]{>{\raggedright\let\newline\\\arraybackslash\hspace{0pt}}X{#1}}
\usepackage{cite}
\usepackage{xcolor}
\usepackage{url}
\usepackage{multirow}
\usepackage{makecell}

\begin{document}
\title{IoT Virtualization: A Survey of Software Definition \& Function Virtualization Techniques\\for Internet of Things}

\author{
	
		Iqbal Alam,
		Kashif~Sharif,~\IEEEmembership{Member,~IEEE},
		Fan~Li,~\IEEEmembership{Member,~IEEE},
		Zohaib Latif,
		Md Monjurul Karim,
		Boubakr Nour,
		Sujit Biswas,
		and Yu~Wang,~\IEEEmembership{Fellow,~IEEE}
	
	\thanks{
		I. Alam, K. Sharif, F. Li, Z. Latif, M. M. Karim, B. Nour, and S. Biswas are with Beijing Institute of Technology, Beijing, China. Email: \{iqbalalam,kashif,fli,z.latif,mkarim, n.boubakr, sujitedu\}@bit.edu.cn.
	}
	\thanks{		
		Y. Wang is with University of North Carolina at Charlotte, NC, USA. Email: yu.wang@uncc.edu
	}
}

\markboth{Copyright Researved.}{}

\maketitle

\begin{abstract}
Internet of Things (IoT) and Network Softwarization are fast becoming core technologies of information systems and network management for next generation Internet. The deployment and applications of IoT ranges from smart cities to urban computing, and from ubiquitous healthcare to tactile Internet. For this reason the physical infrastructure of heterogeneous network systems has become more complicated, and thus requires efficient and dynamic solutions for management, configuration, and flow scheduling. Network softwarization in the form of Software Defined Networks (SDN) and Network Function Virtualization (NFV) has been extensively researched for IoT in recent past. In this article we present a systematic and comprehensive review of virtualization techniques explicitly designed for IoT networks. We have classified the literature into software defined networks designed for IoT, function virtualization for IoT networks, and software defined IoT networks. These categories are further divided into works which present architectural, security, and management solutions. In addition, the paper highlights a number of short term and long term research challenges and open issues related to adoption of software defined Internet of things. 
\end{abstract}


\IEEEpeerreviewmaketitle

\section{Introduction}
The Internet of Things (IoT) \cite{IoT_Iqbal} provides a concept of connectivity of anything from anywhere at anytime so that the interaction of physical objects connected to the network can be done autonomously. 
IoT is closely coupled with sensor technology, because in most of the cases sensors \& actuators are part of a larger IoT network. The use of IoT devices such as laptops, smart-phones, home appliances, industrial systems, ehealth devices, surveillance equipment, precision farming sensors, and other accessories connected to Internet would exceed 45 billion by 2020 \cite{bizanis2016sdn}. These IoT sensors \& actuators may produce large volumes of data. Hence, the need for installing new network access \& core devices will increase. To manage the network devices efficiently, the network hardware resources need to be virtualized.

Virtualization \cite{7387427} is the logical abstraction of the underlying hardware devices within a network, through software implementation. The abstraction decouples the control from hardware, and makes it easier to modify, manage, and upgrade. In recent times, the abstraction has not been limited to hardware only, but rather software embedded into hardware has also been virtualized as independent elements.\par 
Traditional networks are usually rigid and fixed. Heterogeneity, scalability, and interoperability has been major challenges due to rapid growth of Internet. Software Defined Networks (SDN) \cite{8187644} and Network Function Virtualization (NFV) \cite{7243304} are the two main solutions to provide virtualization in communication. SDN physically decouples network control plane from the forwarding plane, and centralizes the decision making for physical forwarding devices. It enables the network control to become directly programmable, and the underlying physical infrastructure to be virtually abstracted for applications \& network services. The OpenFlow (OF) \cite{7929710} protocol is the foundation of communication between SDN-enabled devices. Several benefits of SDN include:~(i)~direct network programmability allowing network managers to configure, manage, optimize, and secure network resources dynamically, (ii)~network-wide traffic flow control \& flow installations, (iii)~network intelligence is logically centralized providing a global view of the network, and (iv)~a vendor independent open standard, simplifying network design \& operations. NFV \cite{7073808,Schaffrath:2009:NVA:1592648.1592659,5183468} is the mechanism of abstracting functions, such as firewall, load balancing, path calculation, etc., from dedicated hardware to virtual environment. The key benefits of NFV  includes replacing dedicated hardware with commodity servers. It enables to host SDN applications like security functions, load balancing, data collection \& analysis, etc., through deployment of on-demand virtual network functions (VNFs). This enables not only enhanced scalability \& elasticity for deploying vendor independent commodities with reduced cost, but also optimizes computing, memory, storage, and networking capacity of network devices. SDN and NFV are not competing technologies, but are complimentary to each other. The key benefits of both technologies are inter-related. NFV can boost SDN towards virtualizing the SDN controller and other network applications in the cloud. Similarly, SDN with its programmable network connectivity can implement traffic engineering decisions taken by VNFs \cite{7350211}.

Use of SDN along with IoT has been studied to some detail. A number of solutions \cite{7218418,6385039,6324377,6838365} have been proposed to address different IoT optimization challenges by using software defined networking. Similarly, Network Functions \cite{8169853} of IoT devices and ecosystem can also be virtualized to make them more agile, robust, \& cost effective. This will reduce the number of physical devices needed, easily segment networks, and enforce security policies on physical devices.\par
It is important to note that in this work we classify virtualization techniques in IoT from three different aspects: (a) SDN-based IoT refers to IoT solutions which use software defined networking for core communication, (b) NFV for IoT includes solutions which virtualize IoT specific functions in the whole ecosystem, and (c) Software Defined - IoT (SDIoT) refers to techniques which not only makes the network layer virtualized but also includes device and application abstraction, virtual security policy implementation, and virtual device configuration and management, etc.

\textbf{Contributions of this work:} In this paper we provide a comprehensive survey of different virtualization solutions designed specifically for IoT. Following is a list of major contributions.
\begin{enumerate}
	\item An overview of general virtualization techniques and their benefit to IoT.
	\item Classification of solutions available in literature.
	\item In-depth survey of SDN-based solutions with respect to architecture, management, and security in IoT.
	\item Detailed analysis of function virtualization techniques in IoT, along with their uses in SDNs for IoT.
	\item Details of software defined Internet of Things and virtualization of different elements in ecosystem.
	\item Major research directions for virtualization in IoT.
\end{enumerate}

\textbf{Structure of Paper:} Table~\ref{tab:Outline} gives an outline of organization of the paper. Section II gives an overview of different virtualization techniques for SDN, control plane, functions, and devices. IoT and details of its working are given in section III. Section IV elaborates the literature classification and other works similar to this article. Section V, VI, and VII provide detailed survey of solutions for SDN-based IoT, function virtualization in IoT, and software defined IoT, respectively, along with analysis and comparisons. Section VIII gives the future research directions, and conclusion is given in section IX.

\begin{table}[!t]
	\centering
	\caption{Outline of this article.}
	\label{tab:Outline}
	\begin{tabularx}{0.99\linewidth}{|>{\hsize=.25\hsize}Z|>{\hsize=1.75\hsize}Y|}\hline
		\textbf{Sections} & \textbf{Details} \\\hline
		
		II & Overview of network virtualization techniques\newline
		\quad$\bullet$\quad Control plane virtualization \& components\newline
		\quad$\bullet$\quad Function virtualization\newline
		\quad$\bullet$\quad Device virtualization\\\hline
		III & Internet of Things\newline
		\quad$\bullet$\quad IoT use case applications\newline
		\quad$\bullet$\quad IoT challenges \& SDN benefits\newline
		\quad$\bullet$\quad IoT stack and protocols\newline
		\quad$\bullet$\quad Sensor networks \& IoT \\\hline
		IV & Motivation \& Related works\\\hline
		V & Software Defined Network based IoT\newline
		\quad Architecture, Security, and Management solutions.\\\hline
		VI & Network Function Virtualization for IoT\newline
		\quad$\bullet$\quad NFV architectures for IoT solution of NFV for IoT\newline
		\quad$\bullet$\quad IoT Management using Virtual Functions\newline
		\quad$\bullet$\quad Security in IoT using of NFV\\\hline
		VII & Software Defined IoT\newline
		\quad$\bullet$\quad Architectural solutions of SDIoT\newline
		\quad$\bullet$\quad IoT Management using SD Frameworks\newline
		\quad$\bullet$\quad Security solution using SDIoT \\\hline
		VIII & Future research directions\\\hline
		IX & Conclusion\\\hline
	\end{tabularx}
\end{table}

\begin{table}[!t]
	\centering
	\caption{List of uncommon abbreviations used in this article.}
	\label{tab:TableIX}
	\begin{tabularx}{0.9\linewidth}{|>{\hsize=.5\hsize}Z|>{\hsize=1.5\hsize}Y|}\hline
		\textbf{Abbreviation} & \textbf{Description} \\\hline
		API & Application Programming Interface\\\hline
		CoAP & Constrained Application Protocol\\\hline
		DDS & Data Distribution Service\\\hline
		DLT & Distributed Ledger Technology\\\hline
		DNP &	Distributed Network Protocol\\\hline
		DPDK &	Data Plane Development Kit\\\hline
		DPI &	Deep Packet Inspection\\\hline
		E/WBI &	East/Westbound Interface\\\hline
		EP &	Entry Point\\\hline
		EPC &	Evolved Packet Core\\\hline
		GMPLS &	General Multi-Protocol Label Switching\\\hline
		HDFS &	Hadoop Distributed File System\\\hline
		HLPSL &	High Level Protocols Specification Language\\\hline
		IPS &	Intrusion Prevention System\\\hline
		LXC &	Linux Container\\\hline
		MANO &	Management and Orchestration\\\hline
		MEC &	Mobile Edge Computing\\\hline
		MIMO &	Multi-input and Multi-output\\\hline
		MINA &	Multi-network Information Architecture\\\hline
		MitM &	Man in the Middle\\\hline
		MNO &	Mobile Network Operator\\\hline
		MPC &	Mobile Packet Core\\\hline
		MQTT &	Message Queuing Telemetry Transport\\\hline
		MVNO &	Mobile Virtual Network Operator\\\hline
		NAC &	Network Access Control\\\hline
		NBI &	Northbound Interface\\\hline
		NFV &	Network Function Virtualization\\\hline
		NFVI &	Network Function Virtualization Interface\\\hline
		NTP &	Network Time Protocol\\\hline
		NV &	Network Virtualization\\\hline
		OF &	Open Flow\\\hline
		ONF &	Open Networking Foundation\\\hline
		ONOS &	Open Network Operating System\\\hline
		OVS &	Open vSwitch\\\hline
		PIGs &	Programmable IoT Gateways\\\hline
		REST &	Representational State Transfer\\\hline
		SBI &	Southbound Interface\\\hline
		SD &	Software Defined\\\hline
		SDNCH &	Software Defined Network Cluster Head\\\hline
		SDNi &  Software Defined Network Interconnection\\\hline
		SDSH &	Software Defined Smart Home\\\hline
		SFC &	Service Function Chaining\\\hline
		SMP &	Security Management Provider\\\hline
		SPF &	Sieve, Process, Forward\\\hline
		TLS &	Transport Layer Security\\\hline
		VF &	Virtual Function\\\hline
		VIM &	Virtual Infrastructure Manager\\\hline
		VNE &	Virtual Network Element\\\hline
		VNF &	Virtual Network Function\\\hline
		VNFM &	Virtual Network Function Manager\\\hline
	\end{tabularx}
\end{table}

\section{Overview of Network Virtualization Techniques}
Network virtualization \cite{8326277} is the mechanism of combining both software \& hardware resources and network functionality into a logically configured single software-based administrative entity. The term \textit{virtual network} refers to the resulting software network entity. In other words, a successful network virtualization would require platform virtualization along with resource virtualization. This is achieved through the Virtualization Layer, which is an additional abstraction layer between network and storage hardware, and the applications running on it. It can be categorized as either an external virtualization, consisting of many networks into a virtual unit, or internal virtualization serving network-like functionality to software containers on a single network server. In the following subsections we elaborate each element of a virtualized network.

\subsection{Control Plane Virtualization} 
Traditionally, a network comprises of hardware devices for connectivity with a dedicated controller built in them. The controller is part of router architecture which instructs switches where to forward packets. Hardware in the physical network devices is managed by the controller. In existing communication network, there is a need for more flexible features from these controllers. An ideal controller can be managed anytime from geographically anywhere in the world. This has opened up the scope for virtualization of the controller, which is implemented through Software Defined Networks (SDNs) \cite{8323847}. The main idea is to separate the control and data plane, i.e. the intelligence of the router/switch is split from the packet-forwarding engine and placed in the control plane. This may be done centrally or in a distributed manner. The SDN controller supports programmability, allowing the underlying infrastructure to be abstracted for applications and network service. Thus, network programmability \cite{7496952} is the process of releasing the network's power in unique ways for more flexible, faster, and intelligent infrastructure that makes the network application-aware. Programmability refers to the ability to enhance network features linking the applications to it and allowing dynamic traffic flow change, providing both network and application-level Quality of Service (QoS).\par 
SDN is a network architecture which can be dynamic, manageable, adaptable, cost-effective, appropriate for high bandwidth requirements, and adapts to dynamic nature of today's applications. It is directly programmable, agile, and centrally manageable. It has the ability to prioritize, de-prioritize or even block specific types of packets with a granular state of control while routing packets in a given network. This process may also be referred to as efficient traffic engineering allowing administrator to use less expensive OpenFlow complaint commodity switches. OF is a communications protocol that allows access to the data plane of a network switch. 
\subsubsection{SDN Architecture} 
\begin{figure}[!t]
	\centering
	\includegraphics[width=0.95\linewidth]{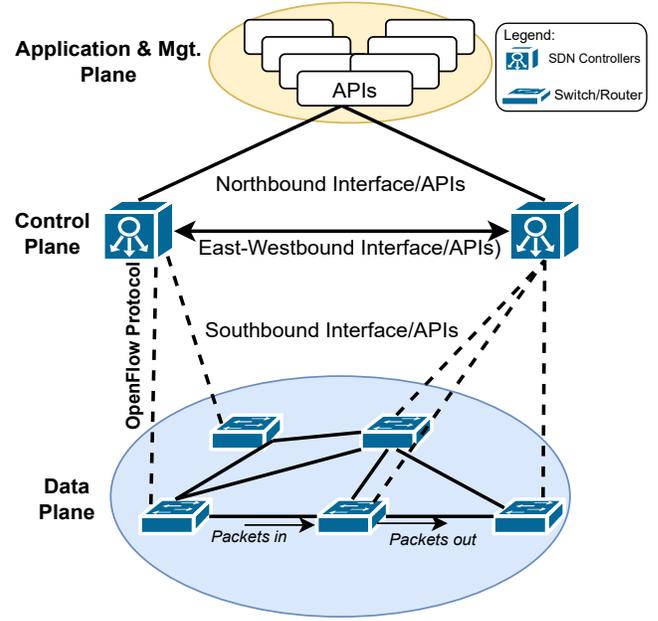}
	\caption{Software Defined Network Architecture.}
	\label{fig:1}
\end{figure}

SDN architecture has three major layers as shown in Figure~\ref{fig:1}. These layers communicate through Application Programing Interfaces (APIs).

\textbf{ Data Plane:} It is also referred as Forwarding Plane, which includes switches, either hardware or software based, connected through a physical medium and perform a set of elementary operations, such as looking up in a table extracting information about incoming packets. These devices have well-defined instruction sets which are used to take actions (forward to port, drop, forward to controller) for incoming packets. These instructions can also be dynamically configured from control plane.

\textbf{Control Plane: }It is a decoupled entity from data plane and is logically a centralized server, also referred as controller, having a global view of the whole network under its control. Based on its global view, it installs forwarding rules on devices in data plane. Some examples of controllers are POX, NOX, OpenDaylight, Floodlight, etc. \cite{Gude:2008:NTO:1384609.1384625,8447647,6838330,ODL,FL,HassasYeganeh:2012:KFE:2342441.2342446}. These controllers can be centralized or distributed \cite{KARAKUS2017279,OKTIAN2017100} in design. 

\textbf{Southbound Interface:} It provides a communication protocol between control plane and data plane. This interface helps controller to program forwarding devices and install flow entries or rules. Some examples of southbound interfaces are \cite{McKeown:2008:OEI:1355734.1355746,doria2010forwarding,Song:2013:PFU:2491185.2491190,pfaff2013open,parniewicz2014design}, but OpenFlow \cite{inferringOFrules_Iqbal} is a widely used protocol in existing SDN implementations.

\textbf{Management Plane:}
Applications designed in management plane can be used to manage and monitor switches in the data plane through the control plane. SDN can be deployed anywhere from enterprise to data centers with the help of management plane, which provides a variety of applications. These applications can be grouped into different categories: Network management and traffic engineering \cite{Jain:2013:BEG:2486001.2486019,6305261,Nakao,Reitblatt:2012:ANU:2342356.2342427,6126682}, Server load balancing \cite{Wang:2011:OSL:1972422.1972438}, Security and network access control \cite{Gember:2012:TSM:2390231.2390233,Gibb2012InitialTO,ETSI,6702549,Nayak:2009:RDA:1592681.1592684,Khurshid:2012:VVN:2342441.2342452,6089085,Jafarian:2012:ORH:2342441.2342467,5689156,5735752,Porras:2012:SEK:2342441.2342466}, Network virtualization \cite{Sharafat:2011:MMV:2018436.2018516,6066002,Gutz:2012:SIS:2342441.2342458,Ferguson:2012:HPS:2342441.2342450,6461196,openstackfoundation}, and Inter-domain routing \cite{Nascimento:2010:QPQ:1851182.1851252,Nascimento:2011:VRS:2002396.2002405,6211892,Caesar:2005:DIR:1251203.1251205,Rothenberg:2012:RRC:2342441.2342445}.

\textbf{Northbound Interface:}
It provides communication between management plane and control plane, where mostly REST API \cite{restapi} is used. There are some controllers (e.g. NOX, PANE, etc.) \cite{Ferguson:2013:PNA:2534169.2486003}, which provide their own northbound APIs and some programming languages (e.g. Frenetic, Procera, etc.) also support them.

\textbf{East/Westbound Interface:}
Scalability and single point of failure are two major challenges in SDN that are resolved by distributed architectures, where multiple controllers work together to attain a global network view. A communication channel is required for these controllers for information sharing. For this purpose East/Westbound Interface (E/WBI) is used, which can interconnect different SDN domains or SDN and traditional network domains. In this context, east refers SDN-to-SDN communication, and west refers to legacy-SDN communication.  

\subsubsection{SDN Functionality Details} The decision making is logically decoupled from the network device, and given to the SDN controller in control plane, which is usually a server running relevant SDN software. OpenFlow protocol (OF) is used by the controller to communicate with a physical or virtual switch in data plane through the Southbound Interface (SBI). Forwarding table is created based on the information provided by the control plane and forwarded to the data plane. Network device (switch) uses this table to decide where to send frames or packets. The routing functionality initiates with the switch encapsulating and forwarding the first packet from a flow to an SDN controller, requesting for addition of a flow to flow table of the switches. When the SDN controller adds the new flow for the switch table, the switch then forwards the incoming packet(s) using the correct port. It is also possible that SDN controller may not add a new flow for the switch in the routing table, and instead enforce the policy to drop the packet during a particular flow, permanently or temporarily, due to security purposes, avoiding Denial-of-Services (DoS) attacks, or for traffic management optimization \cite{CISCO}.

SDN creates dynamic and flexible network architecture, to adapt to the changes in networks requiring rapid deployment. Using centralized control and network automation, SDN also adds more benefits, such as the use of API enabled SDN controllers to execute network commands on multiple IoT devices.

\subsection{Function Virtualization} 
\begin{figure}[!t]
	\centering
  \includegraphics[width=0.95\linewidth]{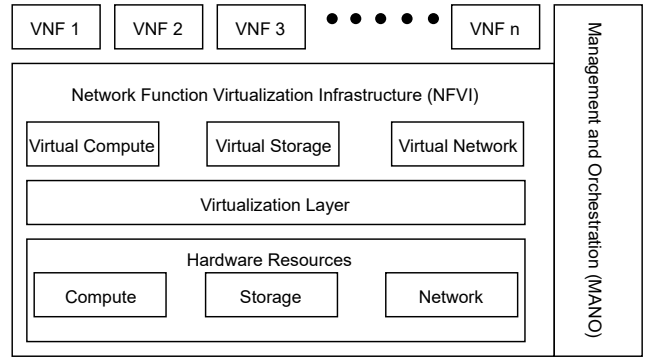}
  \caption{A generic NFV modular structure.}
  \label{fig:2}
\end{figure}

Function Virtualization is implemented through a Network Function Virtualization (NFV) architecture. Figure~\ref{fig:2} shows a generic NFV architecture, which utilizes IT virtualization technologies to virtualize the complete network node functions into series of building blocks to establish connectivity, and to create communication services among them. Its architecture depends on three main components: Virtual Network Function (VNF) \cite{7890085}, Network Function Virtualization Infrastructure (NFVI), and Network Function Virtualization Management and Orchestration architectural framework (NFV-MANO). NFV implements network functions through a piece of software that is configured under NFVI. These network functions tend to be in the form of VNF, which is responsible for handling specific network operations that run on top of the hardware infrastructure. NFVI consists of  both physical and virtual storage, processing, and virtualization software. NFV-MANO architectural framework consists of interfaces and reference points to individual VNFs and NFVI elements.

For example, network function such as firewall, is an instance of plain software, installed inside voluminous switches, storage, and servers, to filter traffic and neutralize vulnerable packets. 
Further benefits include, allowing the relocation and initiation of these nodes from geographically different network locations.

\subsection{Device Virtualization}
Device virtualization is the process of virtualizing a switch in the data plane using certain logical abstractions among its components, or only the functionality to be executed on different operating systems. Virtualization, in a computing platform, tends to hide the physical features from the users, and create an abstract computing platform to define unique rules for switches to comply, which may be referred to as VNFs. The software that controls virtualization is called the control program, also referred to as hypervisor \cite{8404837}. Similarly, Sensor virtualization \cite{doi:sensorvirtualization} provides software abstraction of various external IoT objects, and allows applications to easily utilize various IoT resources through open APIs (e.g. Zeroconf\cite{zeroconf}). Zeroconf or similar APIs allows virtual sensor to transparently discover arbitrary sensor device as virtual switches. 
It is also able to communicate with different applications using a standard communication interface based on UDP/TCP sockets or even HTTP \cite{5416827}. This way, the applications are not required to deal with sensor specific details.

\section{Internet of Things (IoT)} 
Internet of Things~\cite{6722995} is a collection of sensors, actuators, and smart objects, interconnect via the Internet utilizing embedded technology to interact and communicate with the external environment. IoT connectivity and management are two major challenges in its deployment. Usually IoT systems are developed with a specific target and technology.
 
IoT incorporates everything from a small objects to big machines, appliances to building and industries, body sensors to cloud computing. In essence, it has infiltrated every aspect of our lives. \cite{IoT1} estimates that the potential market value of IoT devices and associated technologies will exceed \$14 trillion in the next 10 years. Similarly, major hardware developers (e.g. Apple, Cisco, Samsung, etc.) have made huge investments in different IoT fields.

\subsection{IoT Use Cases} 
\begin{figure}[!t]
	\centering
	\includegraphics[width=0.95\linewidth]{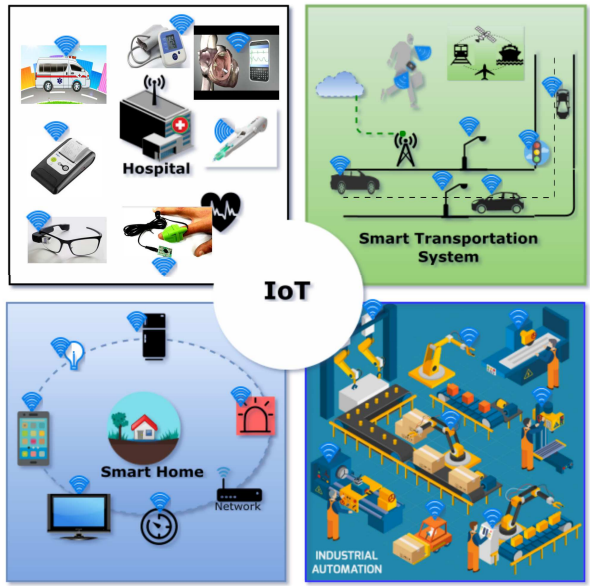}
	\caption{Example IoT ecosystem. Isolated application specific IoT networks may also communicate with each other over the Internet.}
	\label{fig:IoTUsecase2Inscape}
\end{figure}

IoT is playing a significant role in a number of use case applications. Figure~\ref{fig:IoTUsecase2Inscape} shows some examples of IoT ecosystem. The benefits achieved range from small to large scale. Below we briefly introduce some of these use cases, and how they benefit different industries.

\subsubsection{Hospitals \& Healthcare} Application of IoT in both hospital premises and e-health systems is not limited to remote monitoring, but also provides a complete automated healthcare ecosystem. A wide range of IoT devices are used in this process, such as, monitoring cameras, connected inhalers, ingestible sensors, smart insulin delivery devices, smart watch \& wearable sensors/data collectors, connected ambulance, etc.

\subsubsection{Intelligent Transportation Systems} There are various uses of IoT applications in this domain. Sensors are used to retrieve information related to available parking spots for efficient parking management solutions. Smart signboard connected to Internet, can disseminate emergency information alongside roads. 
Asset tracking allows enterprises to easily locate \& monitor vehicular fleets and other mobile assets. Fleet management helps transport companies reduce investment  risks associated to vehicles. It improves efficiency \& productivity, while reducing overall transportation \& management costs. Shipping service uses real time traffic feeds to deliver more packages using efficient algorithms, with lower burden on drivers \& vehicles.
Connected vehicles can better automate many normal driving tasks. Benefits of self-driving cars include accident avoidance, lesser traffic congestion, and other economical efficiencies. Driverless taxis and buses are also a major use case for IoT applications. 
Application of IoT technology in transportation eventually reduces traffic congestion, improves safety, mobility, and productivity.

\subsubsection{Industrial Automation \& Supply Chain} Industrial automation uses artificial intelligence with IoT technology, to automate the supply chain process. Supply chain along with asset tracking optimizes logistics, maintains inventory levels, prevent quality issues, and detect theft. Industry 4.0 production lines are greatly influenced by intelligent manufacturing system, such as smart machines (e.g. multiple smart robots used in car assembling works collaboratively) powered by IoT devices. This results in less human errors, increased speed of production process \& quality of the finished products.

\subsubsection{Smart Homes} IoT in such applications provides a complete intelligent ecosystem for connected devices, ranging from lighting control to security and safety. Usually a smart central hub or gateway is used for human interaction, which in turn controls device automation. These devices can be lined to heating systems, lighting control, appliance monitoring and control, utility usage and optimization, security system, support systems for elderly/disabled, etc.

\subsection{IoT Challenges \& SDN} There are many technological challenges for deploying IoT systems so they can function smoothly. These includes security, connectivity, compatibility \& longevity, standards, and intelligent analysis \& actions. IoT networks are usually large, mobile, and dynamically change their topology \& connectivity. They also have heterogeneous devices which support a range of applications. Hence, challenges like IoT device detection, low power consumption, bandwidth, access control, and data encryption become major concerns for large scale deployment.

SDN ensures reliable connectivity at any given time, based on pre-defined policies. SDN supports customized device configuration enabling efficient packet flows \& optimized routing. It is also a vendor independent platform supporting widely used OF protocol, which mitigates the compatibility standardizing challenges. SDN facilitates device-to-device communication without the intervention of base stations. Heterogeneity is a major concern, especially when billions of mobile IoT devices are connected in a network. NFV plays a significant role in connecting and managing heterogeneous IoT elements. Function virtualization and service chaining mechanisms are the core components to mitigate heterogeneity limitation. Combination of SDN \& NFV supports network programmability, which can improve access control \& bandwidth, data encryption, IoT device detection, low power consumption, etc. for large scale deployment of IoT.

\subsection{IoT Stack and Protocols}
\begin{figure}[!t]
	\centering
	\includegraphics[width=0.95\linewidth]{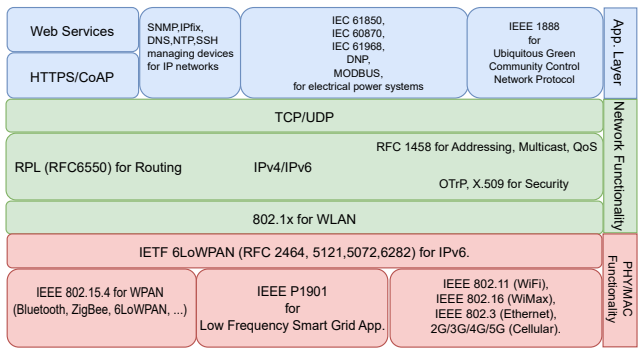}
	\caption{IoT technology stack and protocols.}
	\label{fig:3}
\end{figure}

IoT is applicable in a diverse range of use cases and industries. Its implementation ranges from embedded standalone devices to real-time and mission critical cloud infrastructures. The layered IoT stack shown in Figure~\ref{fig:3}, presents the standards, technologies, and protocols used in such systems. Application Layer specifies all the shared communication protocols and interface medium used by IoT devices. Network Layer specifies communication path over the network (IP address). Physical/Media Access Control (PHY/MAC) Layer specifies communication path between adjacent nodes and data transfer (MAC address). From an SDN perspective, it is very important to understand the technologies used to build IoT networks. It is important to note that SDN does not only install flows for IP packets, but can also be used for radio resource management, security policies, and channel assignment at physical layer, etc.

Various applications fall under the umbrella of IoT, that use different technologies as the main communication enabler~\cite{sheng2013survey, withanage2014comparison}. The most commonly used physical layer technologies are:

\textbf{ZigBee (IEEE 802.15.4)}~\cite{kinney2017ieee,zhou2009wireless}:
Specifies the physical layer and media access control for low-rate wireless personal area networks. It has been designed to run on low-power devices enabling M2M communication. It provides low-power consumption and low duty cycle to maximize battery life. ZigBee can also be used in mesh networks, and supports a large number of devices over long distances with many different topologies, connected all together through multiple pathways.

\textbf{WiFi (IEEE 802.11)}~\cite{deng2015ieee}:
Allows local communication between two or more devices using radio waves. It is the most used technology to connect the Internet gateway to devices. WiFi utilizes both 2.4GHz UHF and 5GHz SHF ISM radio bands. WiFi networks operate in the unlicensed 2.4 radio bands, where the access point and the mobile stations share the same channel and communicate in half duplex mode.

\textbf{Bluetooth \& Bluetooth Low Energy (IEEE 802.15.1)}~\cite{nieminen2015rfc,bisdikian2001overview}:
It is used to transfer data over short distances using  2.4 GHz ISM band and frequency hopping, and up to 3 Mbps data rate with 100m as maximum range. The technology is mostly used to connect user phones and small devices with each other.

\textbf{6LoWPAN}~\cite{shelby20116lowpan,kushalnagar2007ipv6}:
It is a networking technology that combined the Internet Protocol (IPv6) with Low-power Wireless Personal Area Networks (LoWPAN), which is one of the most suitable technologies for IoT deployment. It is a good choice for the smaller devices that are limited in processing and transmission capabilities.

\textbf{5G}~\cite{rappaport2013millimeter}:
The fifth-generation wireless is the newest iteration of cellular technology that is based on the IEEE 802.11ac wireless networking standard in order to speed up the transmission data, reduce the latency. Both LTE and MIMO are used as a foundation in 5G network, as well as network slicing.
	
\subsection{Sensor Networks and IoT}
\textbf{Wireless Sensor Network (WSN)}: It is a distributed and self-organized wireless network that consists of autonomous devices using sensors to observe physical or geographical conditions. According to \cite{niyato2017wireless}, due to the ability to relay messages from one node to another, the area coverage of such networks may differ from a few meters to several kilometers. It is important to note that sensor network and IoT networks are not the same. At best, sensor networks are a subset of IoT ecosystem. They not only differ in deployment, but also in protocols, topologies, use cases, applications, and other technical aspects. A handful of SDN solutions for WSNs have been proposed, but they cannot be directly applied to IoT.

\section{Motivation \& Classification}
In this section we first discuss the existing surveys, and derive the need and motivation for this article. Following it, we present the basic classification and group of literature reviewed.

\subsection{Motivation \& Existing Surveys}
Virtualization, SDN, and IoT have individually attracted attention from the research community. However, there has been very limited effort to review the literature which combines them. Table~\ref{tab:existing_Surveys} lists surveys which have previously been done, and are related to the work in this paper. It is important to note that most of them only target a specific technology. The closest work is \cite{bizanis2016sdn} and \cite{7060643}, which deals with the virtualization in IoT and WSN, respectively. Bizanis~et~al.~\cite{bizanis2016sdn} provide a survey of literature from 2009-2016 and mostly focus on SDN and network virtualization in IoT applications, specific to mobile, cellular and 5G context. It does not cover IoT in-depth nor considers all solutions available. Khan~et~al.~\cite{7060643} focus specifically on WSN and do not collect works on IoT in general.

Some other surveys related to SDN or NFV have also touched IoT in passing. Pan~et~al.~\cite{8089336} focused mainly on IoT application based on future edge cloud and edge computing but the effort is only limited to brief introduction of related challenges and enabling cloud based technologies like SDN and NFV for IoT applications. Akpakwu~et~al.~\cite{8141874} concentrated research on 5G based communication technologies and challenges for IoT applications. But the effort is limited to IoT application usecases for mobile communications. There is only brief introduction of two useful technologies like SDN and NFV to counter IoT management specific issues for future telecommunication system. Cox~et~al.~\cite{8066287} focused research only on SDN state-of-art and challenges with no related solution reviews. Ngu~et~al.~\cite{7582463} presented findings on design of real-time prediction of blood alcohol content using smart-watch sensor data and IoT middleware issues and enabling technologies.

We believe that there is a need to classify and analyze literature, which focuses directly on IoT in terms of different virtualization techniques. Moreover, these virtualization techniques should not be limited to SDNs for IoT, but should also include network function virtualization, network virtualization, and most importantly software defined Internet of Things. 
 
\begin{figure}[!t]
	\centering
	\includegraphics[width=\linewidth]{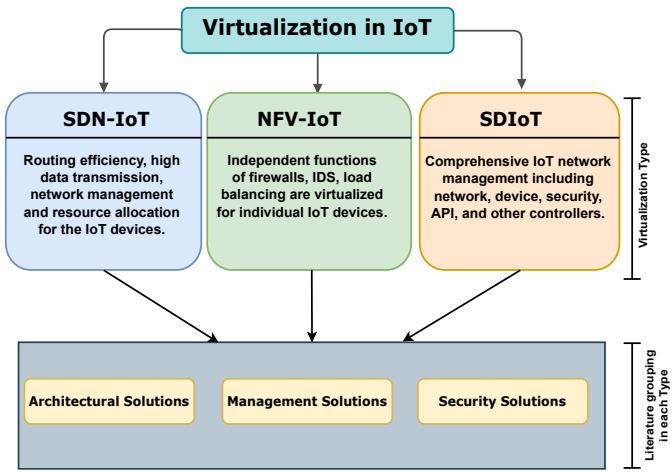}
	\caption{Classification of literature for this article. Literature in three major categories is further grouped into specific types of solutions.}
	\label{fig:Organogram}
\end{figure}
\begin{table*}[!h]
	\centering
	\caption{Summary of related surveys \& contribution of this work.}
	\label{tab:existing_Surveys}
	\setlength\tabcolsep{3.5pt}
	\begin{tabularx}{0.95\linewidth}{|>{\hsize=.6\hsize}Y|>{\hsize=.2\hsize}Z|>{\hsize=1\hsize}Y|>{\hsize=2.2\hsize}Y|}\hline
		\textbf{Survey} & \textbf{Year} & \textbf{Main Focus} &\textbf{Details}\\\hline
		
		Pan et~al. \cite{8089336}	&
		2018	&
		IoT applications based on edge, cloud, \& edge computing.	&
		Brief introduction of challenges \& enabling cloud based technologies for IoT applications: NFV and SDN.\newline[3pt]
		Review covers 2009-2016.\\\hline
		
		Akpakwu et~al. \cite{8141874}	&
		2018	&
		5G for IoT: Communication technologies and challenges.	&
		Limited to IoT application use cases for mobile communications.\newline[3pt]
		Briefly introduces SDN \& NFV technologies.\newline[3pt]
		Review covers 2002-2017.\\\hline
		
		Cox et~al. \cite{8066287}	&
		2017	&
		SDN advancement survey.	&
		Discusses SDN state of art \& challenges.\newline[3pt]
		Brief discussion on SDN-IoT, NFV, and SDIoT.\newline[3pt]
		Review covers 2002-2016.\\\hline
		
		Ngu et~al. \cite{7582463}	&
		2017	&
		IoT Middleware issues and enabling technologies.	&
		Focuses on middleware with limited discussion on virtualization.\newline[3pt]
		Review covers 2003-2016.\\\hline
		
		Bizanis et al. \cite{bizanis2016sdn}	&
		2016	&
		SDN and virtualization for IoT.	&
		Focuses on SDN and NV in IoT applications, specifically in mobile \& cellular context and limited to 5G \& WSN.\newline[3pt]
		Review covers 2009-2016.\\\hline
		
		Khan et al. \cite{7060643}	&
		2016	&
		WSN virtualization.	&
		Limited to detailed discussion about WSN virtualization, state-of-art, and research issues. IoT is not the main focus.\newline[3pt]
		Review covers 2003-2016.\\\hline
		
		This work	&
		2018	&
		IoT virtualization using SDN, NFV, NV, and hybrid SD designs.	&
		Discusses solutions which are specific to IoT.\newline[3pt]
		Literature is covered which utilizes software defined networking (network layer), function virtualization, hypervisors, hybrid NFV and SDN, and software defined Internet of Things.\newline[3pt]
		Review covers all literature till 2018.\\\hline	
	\end{tabularx}	
\end{table*}

\subsection{Classification}
In this work we have categorized the IoT virtualization solutions into three main categories, which are then further divided into 3 types of solutions. The main categories, as shown in Figure~\ref{fig:Organogram} are: 
\begin{itemize}
	\item SDN-based IoT solutions: These solutions only address the virtualization of network layer control (flow management and data transmission).
	\item NFV-IoT solution: These solutions are either in combination of SDN or stand alone but focus on individual functions of IoT ecosystem.
	\item Software Defined IoT solutions: These are more elaborate and provide broader solutions for IoT. 
\end{itemize}
In each category we have grouped the solutions into three types. Some solutions present architectures (with or without implementation), while others are more focused on management of IoT network and devices. The third type are related to security of IoT networks.

\section{Software Defined Network based IoT}
SDN-based IoT is a concept where SDN can facilitate routing efficiency, high data transmission, network management and resource allocation for the IoT devices to meet the growing need of the user demands \cite{Tayyaba:2017:SDN:3102304.3102319}. SDN solutions in IoT environment are expected to resolve traditional network issues \cite{huawei1}, like heterogeneity, interoperability, and scalability among IoT devices, inefficient service deployment (lack of dynamic services), slow adaptation to new services (network upgrade time consumption), and lack of user experience guarantees (minimum bandwidth). To do so, different SDN-based IoT architectures have been proposed in many works until recently. 
Commercial solutions such as AR2500 Series \cite{huawei} agile IoT gateways are also available for deployment. In addition to commercial solutions there are numerous proposals and solutions available in academic literature. We classify them into architectural, security, and management solutions. SDN-based IoT architecture deals with clear separation of concern between services provided in the control plane and the data plane. Control plane specifies the management of network traffic and data plane specifies the mechanisms to forward traffic to desired destination. SDN-based IoT management specifies how the applications on top of the Management Layer interacts with the control plane and the coordination among them. It also allows the admin/analyst to define how the control process is to be governed not only by the SDN controller itself but also by human users. SDN-based IoT security specifies different security parameters for access to network, end-point devices, and other control layer elements. It does this by defining security policies for the complete software defined system.  
\begin{figure}[!t]
	\centering
	\includegraphics[width=0.95\linewidth]{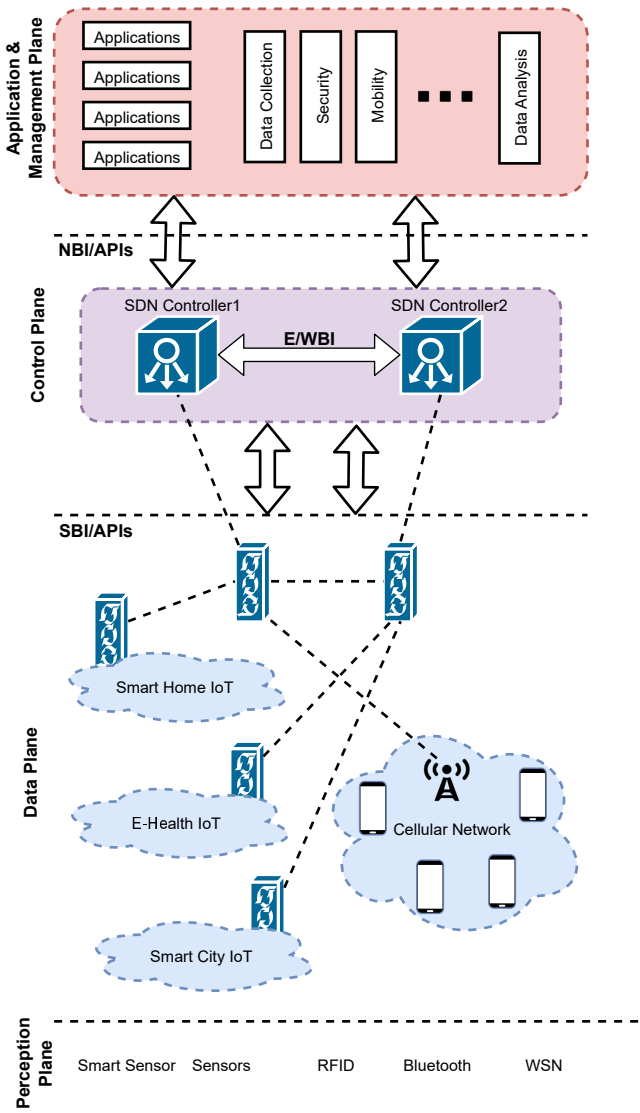}
	\caption{A generic SDN based architecture for IoT.}
	\label{fig:4}
\end{figure}

\subsection{Architecture Solutions} 
Works in \cite{IoT_SDN_OFenabled,IoTEcoSysIqbal,salman2015architecture,SIMECA_Iqbal} propose SDN-based cloud platform approaches for IoT network connectivity, \cite{qin2014software,martinezempowering,li2015general,li2016sdn,SDN_IoT_architecture_with_NFV} propose general SDN-based architectures to facilitate the scalability, heterogeneity, and interoperability among IoT devices or nodes, and \cite{6838330} propose SDN-based control plane platform solutions.  Figure~\ref{fig:4} depicts the SDN-based IoT architecture. It provides a general overview to show the management plane, control plane, data plane, and perception plane. How the IoT sensors would interact with the data and control plane, is discussed in this section through different research solutions.  Table~\ref{tab:TableII} shows these case studies and comparison among different architectures.

\begin{table*}[!t]
	\centering
	\tiny
	\caption{Comparison of different SDN architectures for IoT networks}
	\label{tab:TableII}
	\setlength\tabcolsep{2.5pt}
	\begin{tabularx}{\linewidth}{|>{\hsize=.5\hsize}Y|>{\hsize=0.8\hsize}Y|>{\hsize=1.5\hsize}Y|>{\hsize=.5\hsize}Z|>{\hsize=.6\hsize}Z|>{\hsize=1.65\hsize}Y|>{\hsize=1.45\hsize}Y|}\hline
		
		\textbf {Literature} & \textbf{Objectives}  & \textbf{Solutions}  & \textbf{Control Plane Arch.}  & \textbf{Controller}& \textbf {Benefits} & \textbf {Limitations} \\\hline    
		
		Desai et~al. \cite{IoT_SDN_OFenabled} &
		Heterogeneity, IoT device to cloud comm.	&
		OF-enabled management device	&
		Distributed	&
		NOX, POX, ODL	&
		Proposed an OF-enabled management device, which will make network simpler	&
		Implementation of OF-enabled management device left as future work.\\\hline
		
		Ogrodowczyk et~al. \cite{IoTEcoSysIqbal}	&
		Scalability, QoS, Reliability, Security	&
		SDN-based IoT network application working on top of SDN controller.\newline[4pt] 
		Dynamic creation and management of end-to-end comm. channels among IoT devices and cloud	&
		Distributed	&
		Ryu, OpenFlow	&
		Orchestrated application uses specialized controller for traffic analysis.\newline[4pt]
		IoT device categorization, recognition, \& policy enforcement.\newline[4pt]
		Real-time data collection, visualization, storage and analysis through automated IoT service deployment.	&
		Scenario specific solution (smart cities).\newline[4pt]
		Each sensor is used by a single tenant only, which limits virtualization at device level.\\\hline
		
		Salman et~al. \cite{salman2015architecture}	&
		Single solution for multiple challenges: Scalability, Heterogeneity, QoS, Latency, Reliability, Security.	&
		Centralized SDN control network for IoT with decentralized data management.\newline[4pt]
		Layered model: Works in application, control, network, and device layers.\newline[4pt]
		Implements SD-gateways in the fog with specialized algorithms.	&
		Distributed	&
		SDN controller	&
		Inter-controller communication.\newline[4pt]
		Intelligent fog nodes.\newline[4pt]
		SDN controller uses management protocols (i.e. NetConf and Yang, OF-Config \& extended OF).\newline[4pt]
		Use of unified application for communication.	&
		Architecture only. \newline[2pt]
		Simulation and implementation is left for future work.\\\hline
		
		Nguyen et~al. \cite{SIMECA_Iqbal}	&
		Latency, QoS, Overhead, Mobility.	&
		Services hosted inside edge devices.\newline[4pt]
		Packet header translation.\newline[4pt]
		Separating end point and routing identity.\newline[4pt]
		Lightweight control mechanism.	&
		Distributed	&
		SDN controller	&
		Lightweight solution.\newline[4pt]
		Efficient peer-to-peer service abstraction for IoT devices.\newline[4pt]
		Reduced signaling \& data overhead.\newline[4pt]
		Flexible service deployment \& resource management.	&
		Controller compatibility with the proposed architecture may become an issue.\\\hline
		
		Qin et~al. \cite{qin2014software}	&
		Heterogeneity, Interoperability, Scalability , Security, QoS.	&
		Centralized global view.\newline[4pt]
		Heterogeneous devices with various data formats for information modeling.\newline[4pt]
		Adaptable network state.	&
		Centralized	&
		Layered IoT controller.	&
		Minimized latency and optimized interoperability \& scalability.\newline[4pt]
		Better performance and flow scheduling.	&
		Limited security and tools for resource provisioning or network control.\\\hline
		
		Li et~al. \cite{li2015general}	&
		Heterogeneity, Interoperability, Scalability, Availability, Security.	&
		IoT gateways and SDN switches.\newline[4pt]
		Distributed network OS.	&
		Distributed	&
		SDN controller	&
		Distributed OS providing centralized control.\newline[4pt]
		Global view of the underlying physical distributed network environment.	&
		Architecture only. \newline[4pt] No performance evaluation \& implementation available.\\\hline
		
		Li et~al. \cite{li2016sdn}	&
		Heterogeneity, Interoperability, Latency, Scalability, Reliability, Security.	&
		SDN gateway/router.\newline
		Distributed network OS.	&
		Distributed	&
		POX	&
		Discovering IoT devices from different domains.\newline[2pt]
		Real time evaluation for latency in IoT devices \& sensors, using Raspberry Pi.	&
		No discussion of security mechanisms.\\\hline
		
		Ojo et~al. \cite{SDN_IoT_architecture_with_NFV}	&
		Heterogeneity, Scalability, Mobility.	&
		Replacement of traditional gateway with SDN gateway.	&
		Distributed	&
		ONOS, ODL	&
		Improved network efficiency \& agility.\newline[2pt]
		SDN-enabled gateway.\newline[2pt]
		Intelligent routing protocols \& caching techniques.	&
		Architecture only.\newline[2pt]
		Performance evaluation and implementation left for future work.\\\hline	
	\end{tabularx}
\end{table*}

Desai \textit{et al.} \cite{IoT_SDN_OFenabled} proposes an architecture where IoT device communication with cloud based processing systems is enabled using SDN. The proposed \emph{management device} structure is designed for a number of different applications, such as smart homes, temperature sensors, etc. It also contains application frameworks for system management, communication drivers, Secure Socket Layer (SSL) and media framework libraries, runtime process, \& virtual machines. SSL is used for data encryption. The respective communications driver, depending on the type of IoT device attempting to establish connection with the OpenFlow-enabled management device, uses appropriate libraries for data encryption and decryption. 
The data manger formats the data appropriately for the application layer, and then  forwards to the OpenFlow-switch (OF-switch). The OF-switch works in a traditional manner, and consults the forwarding table for packet processing. Once the data reaches the gateway controller, it negotiates with other gateway controllers to determine the destined location where the data should be processed. The destination may be located in the local domain or cloud domain. In case of cloud domain, the data will be sent to the cloud gateway controller from the local gateway controller and is processed, the output is sent to the respective destination based upon negotiations. The output location can be an IoT device which is attached to an OpenFlow-enabled management device. Since the layered architecture is configured in Linux kernel, it can be considered reliable. The authors suggest that implementation or deployment of OpenFlow-enabled management device is expected to be carried out in the future.

Ogrodowczyk \textit{et al.} \cite{IoTEcoSysIqbal} presents an architecture which contains multiple independent IoT ecosystems connected through cloud using SDN infrastructure. The solution is able to generate a global view of all IoT resources using OF Experimenter extensions for auto-detection. Service provisioning is automated by inserting meta-data into the flow information. The solution also proposes a protocol which interfaces between the cloud orchestrator and the OF controller (Ryu). The IoT services are instantiated inside Linus Containers (LCX), which are virtualized isolated Linux systems (containers) controlled by a single kernel. The orchestrated application uses commercial product NoviFlow \cite{NoviFlow}. The Ryu SDN controller \cite{8370397}, in the proposed architecture, is a customized version, rebuilt from scratch in Python. 
The solution is evaluated in Poznan Smart City use case. The authors demonstrate the \emph{slicing} of a city into different smart spaces, while connected to a single SDN-based platform. The city wide network is an OF enabled infrastructure integrated with cloud resources, capable of hosting multi-tenant cloud applications for IoT devices. The IoT application was tested with sensors like Libelium \cite{libelium} (i.e. IoT gateway to connect any sensor to any cloud platform), and with the Spirent STC \cite{SpirentSTC} ( i.e. a test emulator to analyze complex traffic pattern). For real-time performance evaluation of a smart city and to scale the entire system, further testing is required to validate the feasibility of using vendor independent sensor devices rather than confined to specific sensors.
 
Salman \textit{et al.} \cite{salman2015architecture} proposes an architecture, with layered model, for IoT with decentralized data and centralized control. Authors also discuss IoT challenges like scalability, big data, heterogeneity, and security. The proposed four layered model consists of Application, Control, Network, and Device Layer. The architecture uses unique identifiers in device layer that ensures interoperability, security, and quick address. Software Defined Gateways (SD-Gateways), a virtualized abstraction of a common gateway supporting extended OpenFlow protocol to communicate with the SDN controllers, enforces a Genius algorithm \cite{WILLE2015629} as one of the virtual functions on top of it. They are expected to mitigate the IoT challenges, when applied in the Network Layer. Fog nodes (i.e. SD-Gateways) would bridge the communication between IoT devices and the SDN controllers. The authors leave the implementation of SD-Gateways for future work. Control Layer specifies the network orchestration and computation such as collecting the topology data, defining security rules, implementing scheduling algorithms, and computing the forwarding rules with routing algorithms. However, these algorithms have not been addressed in depth in the paper and may possibly be considered as future research directions. Application Layer reveals the use of software functions based on the information provided by the control layer, which is yet to be implemented.

Many works also discuss the migration of traditional IoT network to SDN. Qin \textit{et al.} \cite{qin2014software} discusses MINA (Multi-network Information Architecture), a centralized architecture for heterogeneous nature of IoT. It attempts to address the interoperability challenges with different heterogeneous devices, and exploits various data formats for modeling information. MINA's objective is to minimize latency and optimize interoperability and scalability to improve QoS. A customized Qualnet \cite{QualNet} simulation platform with SDN features based on OpenFlow-like protocol in IP layer is used. It enables effective resource provisioning in IoT multi-networks environment by using Observe-Analyze-Adopt \cite{6838332} loop. It also defines flow scheduling over multi-hop, and heterogeneous ad-hoc paths. It takes advantage of flow matching using heuristic algorithms (i.e. network calculus and genetic algorithm \cite{Seo:2008:GAS:1363686.1364121}) to examine QoS, considering parameters like jitter, delay, and throughput. Its proposed flow scheduling algorithm proves better compared to the existing ones. However, security and availability for sophisticated tools to enable on-the-fly resource provisioning and network control are left for future research work.

Pedro \textit{et al.} \cite{martinezempowering} aims to enhance IoT network by using SDN Controller with an additional IoT Controller. The proposed model tries to integrate SDN and IoT to resolve heterogeneity issue of objects (i.e. IoT devices). The authors analyze the different types of workloads that IoT elements will push to the network, which determines the structure and modularity of IoT Controller. An IoT Controller acts as a functional block, which receives communication interests by the IoT agent installed into the objects, finds the responder in the network graph, uses routing algorithm to calculate the path, builds the forwarding rules based on the nature of protocols holding the object requested, and finally passes such rules to the SDN Controller to be installed on the forwarders (i.e. SDN switches). The advantage of the proposed architecture is that the IoT Controller tends to reduce the workload of the SDN controller but the limitations still may persist, as the nature of routing algorithm is not described. Latency issue to discover objects may also persist, as the author also state that the IoT Controller may sometimes face protocol compatibility issue and hence some rules may need to be handled by the forwarders.

Li \textit{et al.} \cite{li2016sdn} discusses issues like interoperability from the perspective of devices, data, communication protocols, and re-usability of data generated from IoT devices. Moreover, authors suggest resource utilization, openness and interoperability by using a layered architecture which includes Device Layer (responsible for collecting data), Communication Layer (contains SDN enabled switches and gateways), Computing Layer (having SDN Controller), and Service Layer (which provides services). The IoT devices communicates with the SDN gateway/router through sinks, like Raspberry Pi. The gateway/router then forwards the data to the SDN controller. The SDN controller manipulates the data as per the application requirements located at the service layer. This is done by programming the SDN controller. Limitations of this work include sink and sensing devices, which work independently while only responsible for aggregating and caching the data received from IoT devices. Its architecture lacks security mechanisms and routing algorithms both in SDN controller and IoT Gateway.

Nguyen \textit{et al.} \cite{SIMECA_Iqbal} presents a distributed mobile edge-cloud architecture that enables a new network service abstraction called SDN-based IoT Mobile Edge Cloud Architecture (SIMECA). It aims to improve IoT device communication performance, as compared to the Long Term Evolution/Evolved Packet Core (LTE/EPC) architecture. It realizes the abstraction by lightweight control and data planes that significantly reduces signaling and packet header overhead, while supports seamless mobility. Through evaluations with pre-commercial EPC, SIMECA shows promising improvements in data plane overhead, control plane latency, and end-to-end data plane latency, while coordinating large numbers of IoT devices in cellular networks. PhantomNet \cite{PhantomNet} testbed is used for evaluation purpose. Controller information is not available, which may impact the results as different SDN controllers have different features. The proposed system may not be compatible with all SDN controllers due to SDN controller software, interface, and OS compatibility issues. Other issues like heterogeneity, availability, and scalability may also exist from the perspective of physical devices in the network.
 
Li \textit{et al.} \cite{li2015general} proposes an SDN-based IoT architecture with conceptual virtual functions. It consists of three different layers. Application Layer accommodates IoT servers for various applications and services through APIs, Control Layer accommodates SDN controllers running on distributed OS, and Infrastructure Layer accommodates IoT gateways and SDN switches to enable connections between the SDN controller and IoT devices. It carries different technologies like RFID and sensors using the control plane interface. The benefit of employing distributed OS is that it provides centralized control and global view of underlying physical distributed network environment to process network data forwarding. However, the work presented does not show the performance comparison or real world implementation. The issue of IoT devices is also a concern for this architecture.

Ojo \textit{et al.} \cite{SDN_IoT_architecture_with_NFV} proposes a replacement of traditional IoT gateways, with specialized SDN-enable gateways. These gateways are capable of managing wired \& wireless devices, and claims to be more flexible, efficient, and scalable. Authors also claim that the gateway can perform efficient traffic engineering with intelligent routing protocols and caching techniques across less constrained paths. However, the work is limited in defining intelligent routing algorithms, and performance evaluation or implementation in real time which is considered a future direction.

\textbf{Conclusion:} A number of novel algorithms have been proposed to tackle issue of IoT challenges like heterogeneity, interoperability, latency, security, data manipulation, etc. However, most works only propose the architecture. Real world implementation and experiments are needed to address the performance evaluation. Hence, this is a major research direction for this area. Furthermore, the adaption of existing controllers to IoT is still not completely addressed. Controllers which can seamlessly integrate into access network and can reach devices in the mobile domain, will be necessary to better optimize the IoT ecosystem.

\subsection{Security Solutions} 
Traditionally security mechanisms like firewalls, intrusion detection \& prevention system are deployed at the network edge to prevent external attacks. Such mechanisms are no longer enough, considering the dynamic changes in network topology as a result of IoT nodes joining-in and moving-out. As for internal threats, e.g. if an object is corrupted by virus, other uncorrupted objects may also be exposed to threats.  Hence, the security parameters for both internal and external threats may need to be reconsidered with the flow of technological advancement.\par 

\begin{figure}[!h]
	\centering
	\includegraphics[width=0.95\linewidth]{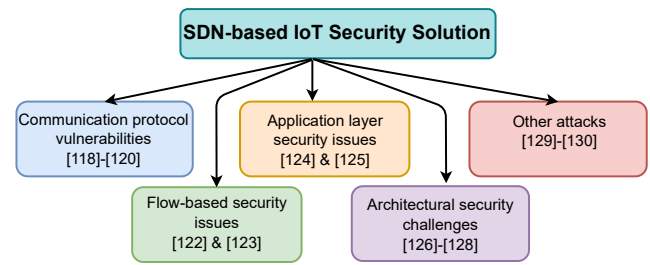}
	\caption{Security solution categorization for SDN-based IoT.}
	\label{fig:SDNIoTSecSol}
\end{figure}

The following literature discusses different proposed solutions for SDN-based IoT security issues. Table~\ref{tab:TableIII} shows comparisons among them. We group these works into different categories, as shown in Figure~\ref{fig:SDNIoTSecSol}: communication protocol related vulnerabilities \cite{xu2017defending,sandor2015resilience,hesham2017simplified}, flow-based security issues \cite{bull2016flow,sivanathan2016low}, application layer security issues \cite{sivaraman2015network} \& \cite{HostBasedIntrusion_Iqbal}, architectural security challenges \cite{flauzac2015sdn,gonzalez2016sdn,7785005}, and other attacks and vulnerabilities which expose the network \cite{li2017securing,chakrabarty2015black}. 

\textbf{Common Protocol Vulnerabilities.} In an SDN environment, the communication between IoT based devices and servers can be blocked by new flow attacks, that contain a significant amount of unmatched packets injected into routing system. This leads to processing of excessive amount of data packets in both data and control plane, and exhaust either the SDN-enabled switch or the controller or both overloaded with intensive new flows, ultimately cutting off the bridge between IoT devices and IoT servers. To solve this issue, Xu \textit{et al.} \cite{xu2017defending} presents a security framework to defend against such suspicious flow attack for IoT centric OpenFlow switches and SDN controllers. The controller acts as a security middle-ware to filter new-flow vulnerabilities, such as DDoS attack, controller-switch communication flooding, and switch flow table flooding, and uses traditional SDN northbound and southbound interfaces to mitigate them. Both simulation and real-time experiments show feasibility to defend against the cyber attacks although calculation process and result filtering technique still need to be improved to implement in a large-scale scenario.\par
Sandor \textit{et al.} \cite{sandor2015resilience} presents an IoT-based hybrid network framework along with a redundant path switching algorithm using SDN's adjustable routing feature, which would protect against DoS attacks. The architecture is hybrid because it includes SDN switches and non-SDN topology segments that contain both types of Entry Point (EPs) and communication edges. By employing SDN switches, the algorithm (i.e. redundant EP switching logic) executes dynamic switching among different EPs. These SDN switches implementing the forwarding plane of the SDN technology are further controlled by the control plane using OpenFlow protocol. These routing rules may also be received from any external entity (e.g. an application to enforce routing policy). Hence, dynamic traffic switching process takes place between two EPs. The authors undertook experiments to measure the performance of the hybrid architecture which exhibited significant reduction in the effect of DoS attack, hence improving the performance and resilience of the IoT systems.\par
Network access control is a security mechanism which limits the access to authorized devices only. Traditional networks use port-based mechanisms defined by 802.1X~\cite{802.1x_rfc}, for its implementation. Hesham \textit{et al.} \cite{hesham2017simplified}, using SDN technology, presents a novel network access control service for IoT sensor networks and M2M communication by replacing the 802.1X standard based software and hardware. The solution also offers adjustment of available bandwidth and predetermined network access policy for each device, to implement authentication and authorization mechanism. This new device should be able to communicate with the OpenDaylight controller via northbound interface. The entire solution consists of four different steps: authenticate clients, authorize clients, flow installations on SDN controller, and deletion of flows on controllers as soon as clients logs out. The solution also follows two separated policy based databases, termed as the User database and the Policy database. The experiment testbed evaluates the system performance for flow installation delay against a varying number of devices and policies. The primary experiment results show some challenges in flow installation, however, the system is able to successfully  authenticate users and register them. 
The results may further be improved by using Apache Cassandra which allows thousands of transactions per second and improves scalability, for policy and authentication database. This will be significantly useful when multiple new devices simultaneously connect to the network (i.e. bootstrapping a new subnet).
However, authors also suggested that the authentication and authorization module could have been wrapped-up inside the SDN controller, which would have improved the performance of the system to a great extent, as less flow installation may mean less time consumed to establish device-to-device connectivity. This can be a possible future direction for research community.

\begin{table*}[]
	\centering
	\caption{Comparison of security solutions using SDN for IoT networks.}
	\label{tab:TableIII}
	\setlength\tabcolsep{2.5pt}
	\begin{tabularx}{\linewidth}{|>{\hsize=.7\hsize}Y|>{\hsize=1.10\hsize}Y|>{\hsize=1.10\hsize}Y|>{\hsize=.7\hsize}Z|>{\hsize=.7\hsize}Z|>{\hsize=2\hsize}Y|>{\hsize=.7\hsize}Z|}\hline   
		
		\textbf {Literature} & \textbf{Objectives}  & \textbf{Vulnerability}  & \textbf{SDN Controller}  & \textbf{Switch Type} & \textbf {Implementation \& Evaluation Details} & \textbf{Operational Layer(s)} \\\hline 
		
		Xu et~al. \cite{xu2017defending}	&
		Detection, Mitigation.	&
		Suspicious flow attack.	&
		ODL	&
		OpenvSwitch	&
		Testbed for attack detection in IoT centric OF switches with OpenDaylight controllers.\newline[4pt]
		Novel packet filtering algorithms implemented in Matlab.	&
		Datalink\\\hline
		
		Sandor et~al. \cite{sandor2015resilience}	&
		Dynamic switching among redundant entry points.	&
		DoS attack	&
		Floodlight	&
		OpenvSwitch	&
		SDN-enabled hybrid network infrastructure, using automatic switching, and advanced routing mechanisms.	&
		Network\\\hline
		
		Hesham et~al. \cite{hesham2017simplified}	&
		Novel network access control mechanism.	&
		Unauthorized access to network devices.	&
		ODL	&
		Pica8 Switch	&
		Testbed with in-band topology (merged control \& data plane) to enable connection between clients \& authentication service.	&
		Datalink\\\hline
		
		Bull et~al. \cite{bull2016flow}	&
		Detection of anomalous behavior in packet flows.	&
		TCP flood attack, DoS attack, ICMP based attack on IoT device.	&
		POX	&
		OF 1.3 Switch	&
		Flow monitoring, periodic checking, and flow installation mechanisms to counter TCP flooding and ICMP attacks.\newline[4pt]
		Mininet based emulation.	&
		Datalink\newline[4pt]
		Network\\\hline
		
		Sivanathan et~al. \cite{sivanathan2016low}	&
		Network level monitoring to detect flow-based anonymous packets.	&
		Self developed two new Python-based emulated attacks.	&
		SDN controller	&
		TP-Link SDN-enabled gateway.	&
		Experimental testbed using C programing.	&
		Datalink\newline[4pt]
		Network\\\hline
		
		Sivaraman et~al. \cite{sivaraman2015network}	&
		Device Monitoring \& Control.	&
		Eavesdropping\newline[3pt]
		Remote access\newline[3pt]
		Privacy\newline[3pt]
		Man in the Middle	&
		Floodlight	&
		OpenvSwitch	&
		Prevention mechanisms for suspicious eavesdropping and packet injection attacks in Smart Home appliances.	&
		Network\\\hline
		
		Nobakht et~al. \cite{HostBasedIntrusion_Iqbal}	&
		Identify and block attacks.	&
		Unauthorized access of smart home devices.	&
		Floodlight	&
		OF Switch	&
		Identify suspicious packet flows \& prevent access to Smart Home IoT devices.	&
		Datalink\\\hline
		
		Flauzac et~al. \cite{flauzac2015sdn}	&
		Distributed routing.\newline[4pt]
		Distributed security rules.	&
		General security issues.	&
		Distributed controller	&
		OF Switch	&
		Multi-SDN domain access control network architecture\newline[4pt]
		Provisioning security for IoT objects (i.e. sensors, smart phone, tablets, etc.).	&
		Datalink\newline[4pt]
		Network\\\hline
		
		Shuhaimi et~al. \cite{7785005}	&
		Reduced hardware usage.\newline[4pt]
		Enhanced security \& privacy.	&
		3rd party applications\newline[4pt]
		Untrusted data\newline[4pt]Privacy	&
		SDN controller	&
		OF Switch	&
		IoT and SDN integrated algorithmic model, to secure attacks from both inside \& outside the domain.&
		Datalink\\\hline
		
		Li et~al. \cite{li2017securing}	&
		Detect Man in the Middle attacks	&
		TLS vulnerabilities	&
		Floodlight	&
		OF 1.3 Switch	&
		Bloom filters based SDN \& extended OF approach to detect MitM attacks emulated using Mininet.	&
		Datalink\\\hline
		
		Chakrabarty et~al. \cite{chakrabarty2015black}	&
		Secure meta-data \& payload within layers.\newline
		Privacy\newline
		Confidentiality\newline
		Integrity\newline
		Authentication	&
		Packet injection\newline[4pt]
		Eavesdropping.	&
		Centralized controller	&
		OpenvSwitch	&
		Payload uses novel encryption mechanism.\newline[4pt]
		Able to mitigate a wide range of passive and active attacks on IoT net.\newline[4pt]
		Uses SDN for routing over multiple topologies.\newline[4pt]
		Node sleep and sync. mechanisms.	&
		Datalink\newline[4pt]
		Network\\\hline
		
	\end{tabularx}
\end{table*}

\textbf{Flow-Based Security.} Data flow related challenges of IoT devices and systems have been described by Bull \textit{et al.} \cite{bull2016flow}, where SDN gateways are used in a distributed structure to monitor data traffic and flow characteristics. The authors propose a method to identify and reduce anomalous behavior, claimed from their previous work in \cite{bull2015pre}, add functionality of packet forwarding/blocking, and enhance QoS by the SDN-based IoT gateways. In this approach, to categorize the network state, source and destination flow statistics are collected from the SDN controller. Additionally, the proposed mechanism executes relevant actions (i.e. permit or block traffic flows) to negotiate with the detected anomalous behaviors. The primary results successfully authenticate the approach by showing a small number of attacks being blocked by using this method, although dynamic traffic analysis and hardware based-testbed experiments are reserved for future works. 

Sivanathan \textit{et al.} \cite{sivanathan2016low} elaborates the differences between flow-based monitoring approaches and packet-based approaches to prevent vulnerabilities in smart-home IoT devices. Based on the flow-level characterization of IoT traffic, the authors present a system containing SDN-enabled gateway with a cloud-based controller to identify malicious IoT activity in the home network. They propose an analysis engine, Security Management Provider (SMP), that communicates with the SDN controller via northbound APIs to recognize trusted IoT devices at low cost. It requests SDN controller to inspect flows selected by it. The SDN controller then configures home gateway with such rules, referred by the analysis engine, to mirror selected traffic flows towards it. It actively inspects the packet in/out of the IoT device with specific headers and also measures the load of selected flows. Traffic analysis is concluded by stopping the traffic mirroring followed by deletion of pertinent rules inside the home gateway. Traffic flows are managed from the cloud-based software, rather than embedded processing unit of home gateway. Internal and external attacks have been demonstrated in an experimental testbed consisting of real IoT devices to prove that the approach can be effective with minimal cost. However, this method is limited to packet content inspection and plain-text password based attack types. Future research may be carried on flow-level monitoring to mitigate other sophisticated security threats.

\textbf{Application Layer Security Issues.} Usage of SDN in IoT for application specific usecase is very important. This also gives rise to security issue. Sivaraman \textit{et al.} \cite{sivaraman2015network} illustrates that a significant amount of IoT based home appliances such as smart bulbs, motion sensors, smoke alarms, and monitoring/analysis devices, lack basic security functions that may have a negative impact on day-to-day activities. The authors argue that security implementation needs to consider various kind of factors like device capabilities, mode of operation, and manufacturer. They propose a prototype, Security Management Provider (SMP), that can control the access to data on devices, by applying dynamic or fixed content-based policies to identify attacks (e.g. eavesdropping, spoofing, etc.) at the network level. SMP exercises configuration control over the ISP network or home router without being directly on the data path. SMP is invoked via API to provide dynamic/on-demand policy, front-end web interface, static policy via web interface, and OpenFlow capabilities. The solution uses FloodLight controller to configure OpenvSwitch (OVS) and Ruby on Rails as security orchestrator and web-GUI developed in Java script A new module is introduced to the FloodLight controller to implement the API for access control, that works as a wrapper to the FloodLight controller firewall, employing access control policies (based on remote IP). These policies are referred by the external SMP entity for a specific home device. Although, the proposed solution has the potential to block threats at the network level, protecting users' privacy still needs to be addressed in detail with regards to the possible exposure of vital personal data. 

Nobakht \textit{et al.} \cite{HostBasedIntrusion_Iqbal} proposes an Intrusion Detection and Mitigation framework (IoT-IDM), providing network-level prevention mechanism against malicious or suspicious ad hoc objects from the external network domain to access Smart Home environment. IoT-IDM users may have enough flexibility to use customized machine learning mechanisms to detect attacks based on learned signature pattern methodology. This framework is realized using SDN technology (i.e. a Java based Floodlight controller) via OpenFlow protocol for remote management purpose and routing efficiency, implemented in real-time using a smart IoT light bulb. However, IoT-IDM works on top of SDN controller, requiring to handle large volume of network traffic. The authors suggest that it is not feasible to use this approach to mitigate the intrusion detection process for all devices, and is applicable to only selected smart home IoT devices.

\textbf{Architectural Security Challenges.} Flauzac \textit{et al.} \cite{flauzac2015sdn} proposes solution which is mainly designed to increase the security of SDN controllers and to solve the scalability issues in multiple IoT-based domains. The work combines wired \& wireless networks, and further extends its solution to ad hoc enabled network and IoT devices like sensors, smart phones, tablets, etc. Each network node acts as a combination of OpenFlow switch and legacy host. Besides, one controller acts as central trusted authority to improve executable security policies while border controllers assist in communication among neighboring IoT domains by establishing communication and exchanging information. However, future work may include the elaboration of management technique of multiple controllers (i.e. security controller, border controller), and inter-SDN controller communication in different layers. It may also include real-time implementation and performance evaluation on how security and border controller may behave and interact among different SDN domains. Security policies may be scrutinized to further enhance access control mechanism.

In a distributed network scenario, Gonzalez \textit{et al.} \cite{gonzalez2016sdn} introduces a proposal that is adequate for an IoT cluster environment, by establishing groups of sensor nodes. OpenFlow and network virtualization technologies have been used for virtual nodes to simulate a distributed cluster based system of 500 devices. Instead of using a traditional approach of the static firewall to block a possible attack, the authors presented an SDN based routing protocol and a dynamic firewall termed as Distributed Smart Firewall that can apply the functionality of an SDN controller. However, the entire framework is not complete as the system is only limited to handle the communications between clusters. Therefore, setting up a dynamic routing protocol along with expanding the simulation to use OpFlex protocol \cite{OpFlexProtocol} functionalities are reserved for future work. Another SDN Controller clustering approach by Shuhaimi \textit{et al.} \cite{7785005}, deals with challenges like availability, heterogeneity, security and privacy in IoT. It also proposes a multi-step novel algorithm, to select SDN Cluster-Head (SDNCH) that works as SDN controller. Its job is not only to manage and control network traffic, but also monitors and prevents the attacks from inside \& outside domains by securing the whole SDNCH domain. It may be considered as a benefit of this proposal, but the work is limited in performance evaluation and implementation. The authors intend to analyze the results from different security attacks such as neighboring attack, black hole, and other related attacks in near future.

\textbf{Miscellaneous Security Challenges.} To detect man in the middle (MitM) attacks in Software-defined IoT-Fog networks, Li \textit{et al.} \cite{li2017securing} proposes a lightweight countermeasure tool. MitM attack is known as one of the common Transport Layer Security (TLS) vulnerabilities \cite{maninthemiddle} and both SDN controller and OpenvSwitch are susceptible to this attack. The authors first demonstrate three different attacks on a simulated environment in Mininet \cite{lantz2010network} using Floodlight controller, and then, by modifying the existing OpenFlow protocol they have proposed a countermeasure to detect these MitM vulnerabilities. The three different attacks are: (i) redirecting flows in the data plane, (ii) exemplifying the attacker's mechanism to collect information from the data plane, and (iii) the attacker's mechanism to infect the controller's view of the  network. The most integral part of this tool has been built inside Floodlight controller, so that modules will be loaded automatically during the initialization of the controller. The experiments have shown a significant improvement in performance and detection accuracy of this method, although the number of false positives remains a concern. Passive attacks may also cause damage to the network. 

Chakrabarty et al. \cite{chakrabarty2015black} proposes Black SDN to secure SDN-based IoT networks. The Black SDN approach encrypts both payload and packet header at the network layer with the implementation of a single SDN controller that has a global view of the existing network. It also helps to communicate with different resource constrained IoT devices through Black packets. This method can mitigate several passive attacks like inference and traffic analysis attacks and also secures meta data which correlates with each packet or frame of an IoT end-to-end device communication, hence improving payload efficiency. The authors demonstrate the working of Black SDN via simulation using various node states and network topologies, and the achieved results proved effective to defend against many passive attacks. Although, Black SDN provides higher level of security than existing protocols but traffic control may become complicated due to the proposed system's increased communication between the SDN controller and the IoT nodes.

\textbf{Conclusion:} A number of security issues and solutions concerning secure efficient packet routing, monitoring, and corrupt packet prevention and access control mechanisms in different operational layers of SDN based IoT network have been discussed. These prevention mechanisms are mostly developed as an external module to cooperate with the SDN controllers. The research community may focus on possibilities to integrate these modules inside the SDN controllers to achieve enhanced scalability. Efforts may be taken to focus on more real-time evaluation against different threat vectors, which can be helpful in determining the status of the solutions.

\begin{table*}[!t]
	\centering
	\caption{Comparison of SDN-based solutions for IoT Management.}
	\label{tab:TableIV}
	\setlength\tabcolsep{2.5pt}
	\begin{tabularx}{\linewidth}{|>{\hsize=.6\hsize}Y|>{\hsize=0.9\hsize}Y|>{\hsize=1.4\hsize}Y|>{\hsize=.7\hsize}Z|>{\hsize=.6\hsize}Z|>{\hsize=1.3\hsize}Y|>{\hsize=1.5\hsize}Y|}\hline      
		
		\textbf {Literature} & \textbf{Objectives}  & \textbf{Solutions}  & \textbf{Control~Plane Architecture}  & \textbf{Controller}& \textbf {Benefits} & \textbf {Limitations} \\\hline 
		
		Hakira et~al. \cite{hakiri2015publish}	&
		Networking, Mobility, Standardization, Security, QoS.	&
		IoT architecture combining SDN with message-based publish/subscribe DDS middle-ware.	&
		Centralized	&
		SDN controller	&
		Filtering \& fusion mechanism for efficient traffic engineering.	&
		Architecture design only.\newline[4pt] No implementation or evaluation.\\\hline
		
		Bera et~al. \cite{bera2016soft}	&
		Real time working, Flexibility, Simplicity.	&
		Device \& Topology management.	&
		Centralized	&
		Customize controller	&
		Application-aware service provisioning, \& improved network performance.	&
		Limited to specific sensor devices.\\\hline
		
		Yiakoumis et~al. \cite{yiakoumis2011slicing}	&
		Scalability, Reduce latency, Efficient load-balancing.	&
		Slicing mechanism using Flowvisor for multiple home networks.	&
		Centralized	&
		NOX	&
		Isolating network traffic \& bandwidth.\newline[4pt]
		Resource sharing.\newline[4pt]
		Cost-effective.	&
		Architecture lacks compatibility with all applications.\newline[4pt]
		Privacy, performance, security, and flexibility may be further improved.\\\hline
		
		Tortonesi et~al. \cite{7543778}	&
		Network load, Storage \& Cost reduction, D2D communication.	&
		Information filtering.\newline[4pt]
		Prioritization using VoI.	&
		Centralized	&
		SDN controller	&
		Reduced load by information filtering.	&
		Distributed and disruption tolerant architectures.\newline[4pt]
		Efficient information processing functions.\\\hline
		
		Fichera et~al. \cite{fichera2017experimenting}	&
		Heterogeneity, Scalability.	&
		Management of data path across IoT, cloud, and edge network.	&
		Distributed	&
		ONOS	&
		Congestion recovery with reliable data delivery.	&
		Redirection of flows may create delays for time-sensitive mice flows.\\\hline
		
	\end{tabularx}
\end{table*}

\subsection{Management Solutions}
At the existing scale of deployed networks, it is almost impossible to manually configure remote devices. IoT requires that network providers are able to configure and reconfigure devices across the network from a centralized management point. However, this requires the right technology to automate the whole management process. SDN is able to facilitate advanced mechanisms to configure and manage devices (e.g. SDN-enabled switch) across variety of different types of networks. This section discusses different proposed SDN-based IoT management solutions. Table~\ref{tab:TableIV} shows management based comparison of existing SDN-based IoT literature.

Hakiri \textit{et al.} \cite{hakiri2015publish} discusses five key network related challenges of IoT, such as current standardization efforts, mobility management, recurring distributed systems issues, communication protocols, and security \& privacy. They outline an IoT architecture that combines SDN with message-based publish/subscribe Data Distribution Service (DDS) middle-ware to solve variety of issues like networking, mobility, standardization, and QoS (Quality of Service) support. In this framework, smart devices are linked with SDN-based IoT gateways to communicate with SDN forwarding devices. Furthermore, an SDN controller connects to the forwarding devices using southbound APIs allowing asynchronous, anonymous, and many-to-many communication semantics. Within a domain, DDS can provide discovery and communication service between different heterogeneous IoT devices and the controller itself. DDS is utilized in local network whereas SDN is responsible for allowing the connection outside of a local network. A novel SDN-enabled gateway is proposed for smooth handover migration between smart IoT devices in a Wide Area Network (WAN). Future work may be on developing the algorithms for the proposed DDS, defining various communication patterns (i.e. transactional queues for request/response interaction, delivery response, event-based interaction) to publish/subscribe data. Algorithms may also be developed on how to differentiate and prioritize traffic packets.

Bera \textit{et al.} \cite{bera2016soft} proposes leveraging of IoT related application-aware service in Wireless Sensor Network environment. They present an architecture named Soft-WSN that is based on the centralized provisioning of SDN controller. The architecture is divided into three layers: application, control, and infrastructure layer. Application layer generates application specific request to be sent to the SDN controller in the control layer. Control layer has SDN controllers to configure the SDN-enables switches. Control layer has two important entities to assist with policy management. First one is device manager, which deals with device specific control tasks such as scheduling the sensing tasks, sensing delay task, and active-sleep management. Second is topology manager, which deals with network topology control mechanism while focusing on the network connectivity management and forwarding rules. Hence, the topology management system can identify every single node and therefore, it can assist SDN controller to provision according to given configuration policies. The proposed system will be effective for several IoT applications. For example, environment monitoring, traffic monitoring, smart home from both topology and device management perspective. From the experiment results, authors show that Soft-WSN provides better data delivery rate, energy efficiency, and traffic overhead than traditional WSN. However, this method has some compatibility issues with other radio technologies like Bluetooth, and controller placement problem may arise under minimized network delay and the overhead of control messages.

Slicing techniques has always played a key role towards securing and managing a complex network. Network slicing is an effective and powerful virtualization capability. It allows creation of multiple logical networks built on top of a common physical infrastructure. This helps in addressing the efficiency, cost, and flexibility requirement of future networks. Technologies like SDN (through network programmability) and virtualization are the means to realize network slicing. Slices may be optimized in many ways including bandwidth and latency requirements. Usually slices remain isolated from each other in the control and user planes. From the user's perspective they only visualize a single network, regardless of the fact that it may physically be a portion of a layered network. Yiakoumis \textit{et al.} \cite{yiakoumis2011slicing} proposes a prototype where multiple home networks can be sliced and a trustworthy third party can manage whole network using different slicing techniques. Similarly, resources can be shared among multiple service providers to reduce the cost. Authors use FlowVisor \cite{sherwood2009flowvisor} for slicing mechanism in OpenFlow networks, providing bandwidth and traffic isolation. SNMP protocol is used to configure the wireless access-points (such as WiFi, SSID, queues, encryption, etc.), and also to inter-operate with firewalls and NATs in smart home environment (i.e. UDP-in-TCP tunneling). The OpenFlow controller (i.e. NOX) independently controls and manages programmatic control of a slice. It also defines the forwarding logic for a switch (in data plane) to operate. The experiment analyzed Flowvisor which enabled high scalability with low latency, showing efficient load-balancing feedbacks. Future work may include extending OpenFlow protocol to virtualize multiple resources in the proposed scheme: Virtual Device Configuration, Virtual Links, and Virtual Address Space. Moreover, improvement in trade-offs among privacy, performance, security, and flexibility can be future research directions.

SDN technology allows installation and management of communications and computational resources to develop and deploy IoT applications. Sieve, Process, Forward (SPF) by Tortonesi \textit{et al.} \cite{7543778}, is an extended SDN architecture of Open Networking Foundation (ONF). The authors use SPF for information processing, and replacement of data plane with Dissemination plane (i.e. data forwarding plane), and uses a novel SPF-Controller, with Programmable IoT Gateways (PIGs). It uses a solution of data processing (i.e. audio/video analysis, IoT device discovery, tracking \& counting) at the edge of the network rather than in cloud, which reduces high bandwidth usage. SPF architecture has three stakeholders: administrators, service providers, and users. Administrators deploy, run, and operate SPF controllers along with PIGs, allowing the service providers to use it. Service providers develop, deploy, and manage IoT applications. Users may utilize the SPF applications available to them by installing its client app on their smart devices. Moreover, critical information is prioritized by ranking objects (i.e. IoT devices) using Value of Information (VoI) metrics. Future work may include extension of SPF-Controller incorporating interesting functionalities which can extract informations from Twitter, Facebook, Wechat, etc. through mobile devices of customers for data analysis purpose. Moreover, improving information filtering mechanisms of PIGs, utilizing both semantic methodologies and complex event processing, can be done. 

A real-time 5G Operating Platform proposed by Fichera \textit{et al.} \cite{fichera2017experimenting}, is able to manage the heterogeneity and scalability of a network. A testbed has been presented in this work for exploiting SDN management capabilities to provide data delivery paths across different network domains under 5G communication. The experiment divides the testbed into IoT-based, cloud-based, and edge networks. To consolidate communication between these environments, an SDN Orchestrator is designed as an application, running on top of an ONOS controller. It is implemented within IoT domains and cloud environment, exploiting network programmability among sensors and Virtualized Functions (VFs), respectively. The real-time 5G operating platform is interlinked to resource infrastructure managers/controllers (i.e. Cloud controller, SDN controller, IoT device manager), lying underneath all the hardware resources (e.g. SDN switch, gateways). Service Orchestrator deals with cloud, SDN, and IoT Orchestrator followed by the respective resource infrastructure manager/controllers. 
SDN controller configures routing policies on flow tables for SDN switch, to enable end-to-end IoT device communication. An SDN Orchestrator is able to recover congestion events (e.g. service outages or degradation events) through the traversable path that has been redirected towards those switches or links that rely on constant monitoring of throughput data. Cloud, SDN, and IoT Orchestrator(s) rely on Service Orchestrator. It is responsible for invocation of services through intent-based interfaces and infrastructure service abstractions. Experimented results show that redirected operation took less time, although, packet dropping at congested switch may tend to degrade the real-time assured services of the proposed scheme.

\textbf{Conclusion:} Most of the SDN-based management solutions available deal with data distribution data services, topology management, home network slicing, and resource management. Some of the directions which can be further explored are synchronization \& compatibility of IoT devices. APIs for such services can improve heterogeneity in the IoT ecosystem.

\section{Network Function Virtualization for IoT}
\begin{figure}[!t]
	\centering
	\includegraphics[width=0.95\linewidth]{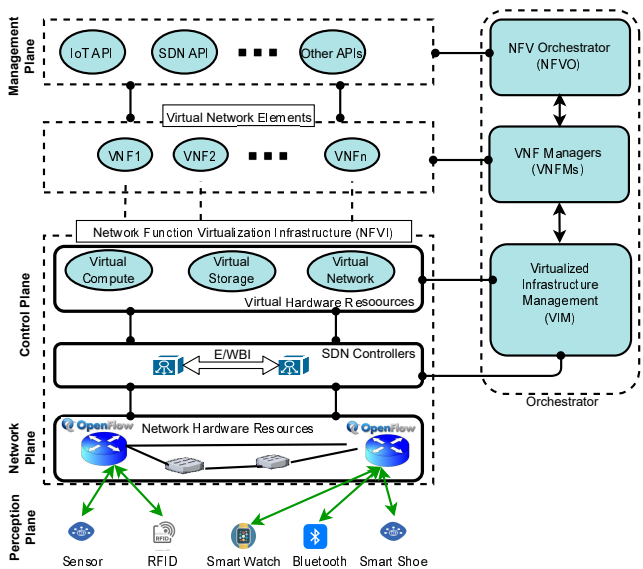}
	\caption{A General SDN-IoT Architecture with NFV.}
	\label{fig:5}
\end{figure}

Network Function Virtualization and SDN are complimentary technologies. They do not require or are dependent on each other, but rather improve and facilitate each other's working. NFV provides a collection of virtual applications referred to as Virtual Network Functions (VNFs). These can include processes for deep packet inspection (DPI), routing, security, and traffic management, which can be combined to provide network services specialized for IoT. A hybrid SDN/NFV architecture for IoT, given in Figure~\ref{fig:5}, shows a general interaction of SDN and NFV to provision reliable communication and to facilitate IoT platforms. The architecture is composed of Network Function Virtualization Infrastructure (NFVI), Virtual Network Functions (VNFs), and Management and Orchestration (MANO) plane, leveraging each other to achieve sustainable network virtualization, with uninterrupted network connectivity, and enforcing efficient packet flow rules by the SDN controller. Different components of this architecture are detailed below:
	
\textbf{Network Function Virtualization Infrastructure}: It consists of all of the networking hardware and software resources required to connect and support carrier network. These resources include operating systems, hypervisors, servers, virtual switches, Virtual Machines (VMs), Virtual Infrastructure Managers (VIMs), and any other virtual and physical assets enabling NFV.  

\textbf{Virtual Network Functions}: VNF focuses on network service optimization. It is responsible for managing specific network function that executes on one or multiple VMs. These VMs work on top of physical hardware resources (i.e. switches, router, etc.). Virtual function for routing, firewall, load balancing, Intrusion Prevention System (IPS), etc. defining unified policy for virtualized hardware resources is adopted into a single VNF. In this way, multiple VNFs may be linked together. This linking can form a service chain managed by VNF manager and VIM, respectively.

\textbf{Management and Orchestration Plane}: MANO facilitates connection of services of different modules of NFVI, VNF, and APIs from the Management Plane, and coordinates with the respective subcomponents in MANO plane.

\begin{itemize}
	\item{NFV Orchestrator:} NFVO works concurrently with VNFM and VIM, standardizing the functions of virtual networking and enhancing the interoperability of IoT devices. It binds together different functions like service orchestration, coordinating, authorizing, releasing, and engaging NFVI resources, to build an end-to-end resource coordinated service in a dispersed NFV environment. 
	\item{VNF Manager:} All VNF instances are associated with VNFM. Its operations include initiation, scaling, updating and/or upgrading, and termination of VNFs. 	
	\item{Virtual Infrastructure Manager:} Network hardware resources like IoT gateways, SDNvSwitches, routers, etc., are abstracted through the virtualization layer using VIM. It keeps allocation inventory of virtual and hardware resources, and manages VNF forwarding graphs, security group policies, hardware resources in a multi-domain environment or optimize them for a specific NFVI environment.
\end{itemize}
	
The rest of the section presents architectural, management, and security solutions of NFV for IoT.
\begin{table*}[]	
	\centering
	\caption{Network Function Virtualization Solutions for IoT networks.}  
	\label{tab:TableV}
	\tiny
	\setlength\tabcolsep{2.5pt}
	\begin{tabularx}{\linewidth}{|>{\hsize=.45\hsize}Y|>{\hsize=1.25\hsize}Y|>{\hsize=1.35\hsize}Y|>{\hsize=.5\hsize}Z|>{\hsize=.75\hsize}Y|>{\hsize=1.7\hsize}Y|>{\hsize=1\hsize}Y|}\hline      
		
		\textbf {Literature} & \textbf{Objective(s)}  & \textbf{Solution(s)}  & \textbf{Control Plane Arch.}  & \textbf{Controller \& Switch} & \textbf {Implementation, Evaluation, Benefits} & \textbf {Limitation(s)}\\\hline 
		
		Jie Li et~al. \cite{li2015general} \newline[2pt]$\bigcirc$ $\bigtriangleup$	&
		Routing, Access control, Security, Traffic control, Virtualization.	&
		SDN-based IoT framework with NFV implementation.	&
		Centralized	&
		SDN controller \& switches with IoT gateways.	&
		Distributed OS.\newline
		Performance, scalability, availability, and security are enhanced due to virtualization.	&
		Limited to the study of general SDN \& NFV architecture.\\\hline
		
		Ojo et~al. \cite{SDN_IoT_architecture_with_NFV} \newline[4pt] $\bigcirc$ $\bigtriangleup$	&
		Interoperability, Device discovery, Scalability, Security, Efficiency and management flexibility.\newline[4pt]
		Application specific requirement provisioning.	&
		An SDN-IoT architecture with NFV implementation.	&
		-	&
		SDN controller\newline[4pt]
		Virtualized IoT gateways	&
		Enhanced performance \& management of hardware, software, \& virtual resources.\newline[4pt]
		Device discovery with enhanced connectivity.	&
		Scalability issues still persists due to overloading of data traffic.\\\hline
		
		Batalle et~al. \cite{batalle2013implementation} \newline[4pt]$\bigstar$	&
		Efficient routing.\newline[4pt]
		Cost effective deployment.	&
		Resolves CAPEX issues in IoT.	&
		Centralized	&
		SDN controller	&
		Efficient inter-domain routing.\newline[4pt]
		Less connected \& deployed devices, hence cost-effective.	&
		Latency\\\hline
		
		Du et~al. \cite{du2016context} \newline[4pt] $\bigcirc$ $\bigtriangleup$	&
		Security \& privacy.\newline[4pt]
		Cost effective \& cheaper IoT \& MVNO.\newline[4pt]
		Value-added services for MVNOs.\newline[4pt]
		Multi-MVNO networks.	&
		Context-aware processing/forwarding of IoT traffic.\newline[4pt]
		Contextual info. recvd. from sensor-layer and application-layer.\newline[4pt]
		Bridges gap between IP \& IoT network, using SDN and NFV.	&
		Centralized	&
		Central service controller.\newline[4pt]
		IoT gateways.\newline[4pt]
		MVNO switch.	&
		IoT framework deployable in current Internet.\newline[4pt]
		Cost effective business model for MVNO use in IoT.\newline[4pt]
		Programmable MVNO IoT gateways (using Edison \cite{Edison} board).\newline[4pt]
		Trailer-slicing for IoT networks.	&
		Proposed arch. may not become a unified IoT platform.\\\hline
		
		Balon et~al. \cite{balon2012mobile} \newline[2pt] $\bigstar$ $\bigtriangleup$	&
		Costs effective.\newline[4pt]
		Study cost-benefit analysis of MVNO, MNO, \& security measures.&
		MVNO based arch. evolution and economic stakes.	&
		-	&
		-	&
		Business model suggesting sharing of info among operators to reduce cost.	&
		Limited to business model.\newline[4pt]
		No implementation.\\\hline
		
		Vilalta et~al. \cite{vilalta2016end} \newline[4pt] $\bigcirc$ $\bigstar$	&
		Low cost IoT.\newline[4pt]
		Enhanced scalability \& interoperability.	&
		An SDN/NFV-enabled edge node, which orchestrates end-to-end SDN IoT services.	&
		Distributed	&
		SDN controller\newline[4pt]
		IoT gateways\newline[4pt]
		OF-switches	&
		ODL \& OpenStack Nova/Havanna service controller.\newline[4pt]
		GMPLS controlled optical network.\newline[4pt]
		Multi domain network architecture.\newline[4pt]
		Optimized packet response time.	&
		Not a unified IoT platform.\\\hline
		
		Salman et~al. \cite{salman2015edge} \newline[4pt] $\bigcirc$ $\bigtriangleup$	&
		High level management capabilities.\newline[4pt]
		Low latency \& Heterogeneity.\newline[4pt]
		Mobility using fog computing.	&
		Edge computing enabling the IoT.	&
		Distributed	&
		ODL, Onix \& ONOS controllers\newline[4pt]
		SD Fog gateways\newline[4pt]
		SD-MEC WHAT IS MEC HERE?\newline[4pt]
		OF-switches	&
		Supports multiple identification and comm. technologies.\newline[4pt]
		Multiple SD fog gateways ensure interoperability.\newline[4pt]
		Centralization leading to security enhancement to some extent.HOW IS IT CENTRALIZED?\newline[4pt]
		Ensures fine-grained flow services using FlowVisor or OpenVirtex.	&
		Scalability.\newline[4pt]
		Infrastructure enhancements exposed to third party causing security vulnerabilities.\\\hline
		
		Maksymuk et~al. \cite{maksymyuk2017iot} \newline[4pt] $\bigcirc$ $\bigstar$	&
		Scalability.\newline[4pt]
		Efficient interoperability \& traffic engineering.	&
		Framework for monitor IoT devices in SD 5G networks.	&
		Centralized	&
		SDN controllers	&
		Architecture based on independent IoT system \& shared by multiple MNO.\newline[4pt]
		Upgraded MNO parameters to include carrier freq., node velocity, cell ID, etc.\newline[4pt]
		use of MQTT to customize monitoring system.\newline[4pt]
		Low traffic overhead.	&
		Not a unified IoT platform.\newline[4pt]
		Third party service involvement may cause security threats.\\\hline
		
		Zhang et~al. \cite{Zhang:2016:OPH:2940147.2940155} \newline[4pt] $\bigstar$	&
		Efficiency \& Scalability.	&
		Dynamic manipulation of packets using NFs in docker container.	&
		Distributed	&
		SDN controller	&
		NF-Lib facilitates fast deployment of NFs.\newline[4pt]
		Improved scalability.	&
		Third party library functions may pose to security threats.\\\hline
		
		Massonet et~al. \cite{8114476} \newline[4pt] $\bigtriangleup$	&
		Enhance security.	&
		NFV/SFC approach.	WHAT IS SFC?	&
		Distributed	&
		SDN controller	&
		Integrated federated agent in IoT network controller \& gateway.\newline[4pt]
		Security VNF within the federated IoT-cloud.	&
		Limited to architecture design only.\newline[4pt]
		Implementation and evaluation left for future work.\\\hline
		
		Al-Shaboti et~al. \cite{8432333} \newline[4pt] $\bigtriangleup$	&
		Enhanced security \& latency.	&
		IPv4 NFV-based ARP server providing security against ARP spoofing \& network scanning.	&
		Centralized	&
		Ryu controller	&
		NFV dispatcher for packet inspection.\newline[4pt]
		Secure ARP operations through NFV-based ARP server.\newline[4pt]
		Not dependent on mapping between the host \& the port.\newline[4pt]
		Both WiFi \& Ethernet port is usable simultaneously.\newline[4pt]
		Reduces packet processing delay.&
		Focus only ARP attacks.\newline[4pt]
		IPv6 for IoT is not considered.\\\hline
		
		\multicolumn{6}{>{\hsize=\dimexpr7\hsize+7\tabcolsep+\arrayrulewidth\relax}X}{$\bigcirc$ $\bigstar$ and $\bigtriangleup$ represent architecture, management and security based solutions respectively.}\\
		
	\end{tabularx}
\end{table*}

\subsection{Architectural solutions of NFV for IoT}
In this sub-section we review architectural solutions proposed in literature for function virtualization in IoT environments. Many of these solutions are hybrid SDN/NFV solutions, which take advantage of each other's capabilities.\par
 
Li \textit{et al.} \cite{li2015general} proposes one such architecture following a top-down approach. It is divided into application layer (e.g. services like Operation Support System/Business Support System), control layer (i.e. SDN controller with distributed operating system), and infrastructure layer (i.e. IoT switches and gateways). The primary objective is to employ SD \& NFV to meet the IoT challenges, such as heterogeneity, scalability, security, and interoperability. The proposed SDN-based IoT architecture with NFV implementation, can provide a centralized control, and virtualize different IoT services in healthcare, transport, education, etc. The proposal only discusses architectural details of how these services may be realized, and leaves out implementation of methodologies. The authors intend to study the organization and components of each part of SDN/NFV-based IoT framework as a future direction.

Ojo \textit{et al.} \cite{SDN_IoT_architecture_with_NFV} presents an IoT framework based on virtualized elements in an SDN-enabled system. They utilize VNFs for a number of purposes, which are deployed on SDN/NFV edge nodes. By using these edge devices the framework is able to provide services such as, rich user context (location information), low latency, high bandwidth guarantees, and rapid IoT device deployment.
The MANO plane orchestrates control of the network infrastructure and the different network functions through respective managers. It also interacts with the management plane applications to obtain policy and configuration information, and with SDN controllers for communication and network services. The overall architecture is quite similar to the one depicted in Figure~\ref{fig:5}. The SDN elements are logically separate from the NFV layers, and some of the functions of SDN are performed in the NFVI. NFV can also be used to relocate some of the IoT gateway functionality into virtual gateways, which will allow greater scalability, easier mobility management, and faster deployment. Although, theoretically the models proposed in this work are sound, there is no implementation or evaluation available to realize the system. Authors have left it as future direct, which can be taken by the research community to integrate SDN and NFV for IoT.

Du \textit{et al.} \cite{du2016context} focuses on prototyping context-aware forwarding/processing mechanism that can manage IoT traffic depending on contextual information. These contextual information is distributed from both sensor-layer and application-layer to mitigate the challenges of IP-based network and IoT network. These issues are related to scalability, discoverability, security, and reliability, mostly due to computation and battery power limitations. The proposal focuses on software-defined data plane defining novel services for Mobile Virtual Network Operators (MVNOs), which offers network services to customers at low prices by means of obtaining network services from Mobile Network Operators (MNOs), without requiring to have their own wireless network infrastructure. The architecture incorporates programmable MVNO switch on multi-core processors and IoT gateways with Edison \cite{Edison} board. The authors focused at high security and privacy mechanism, performance optimization, and value added service in the IoT-MVNO domain. The MVNO switch collects data from the sensors (e.g. smart watches, wearable glasses) via IoT gateways and sends it to the logical service controller for data processing. Several isolated MVNO networks are associated with different applications to work simultaneously. The architecture uses OpenFlow protocol to communicate between the MVNO switch and IoT application through southbound interface. The MVNO switch is built on FLARE \cite{FLARE} testbed equipped with multi-core network processor. IoT Gateway software ensures trailer slicing on FLARE platform, serving functionalities like IoT device discovery and connectivity, data collection and encapsulation, and context-aware packet forwarding/processing. The simulation output shows significantly high rate of data transmission with low bandwidth and efficient routing. The architecture is also realized as an effective business model for IoT application based on MVNO network to make it highly cost-effective. Future plans include a contextual IoT trailer architecture for a unified IoT platform on top of current Internet protocols.
 
Balon \textit{et al.} \cite{balon2012mobile} proposes a model for robust security and network performance management. They show a usecase to build a private virtualized MVNO, which can easily be expanded and scaled for high volume traffic and number of user. They also discuss the different components and enablers of MVNO networks and provide a cost-benefit analysis of using MVNO. However, the paper only discusses architecture and market analysis, but does not give details on implementation using the MNO services. 

Vilalta \textit{et al.} \cite{vilalta2016end} proposes an SDN-based NFV edge node. The proposed edge node adopts OpenFlow-enabled switch, controlled by edge SDN controller. It also provides storage resources, and computing services via edge cloud/fog controller. The OpenStack Nova handles the NFV framework through the Cloud/Fog Network orchestrator, which has two different orchestrators running below it: (i) Cloud/Fog orchestrator, which deals with the edge cloud \& metro controllers, and (ii) Multi-domain SDN orchestrator, which deals with edge SDN \& DC SDN controllers. This entire orchestration consolidates NFV and SDN together to provide seamless network connectivity between deployed VMs to virtual switch at the edge node or in DC. The IoT gateway acts as the client which requests computing and storage services to the SDN/NFV edge node. Multi-domain SDN orchestrator simulates OpenDayLight and OpenStack Nova to provide end-to-end network services. Eventually, data from IoT gateway flows to the processing resources, which are located in the proposed SDN/NFV edge node. The proposed approach is only limited to the edge nodes and DC. The packet response time is considerably low between the IoT Gateway and edge node VM or core DC VM, which optimized the edge resource usage.

Another similar approach towards edge networking is done by Salman \textit{et al.} \cite{salman2015edge} that presents a fog computing architecture termed as Software-Defined Mobile Edge Computing (SD-MEC) for integrating Mobile Edge Computing (MEC) with IoT, SDN, and NFV. SD-MEC is a four-layer architecture that includes an application layer, a control layer, a device layer and a network layer. All of these layers initiate different tasks for the orchestration of the proposed fog services. In this framework, Software Defined Function (SDF) Gateway plays an essential role. It acts as an inter-operator between the various communication protocols and heterogeneous networks, presenting high management capability, rendered from the SDN features, and also offering heterogeneity abstraction, low latency, and mobility support from the fog devices. Applying the NFV features further facilitates management at network level required in the MEC platforms. However, this work only gives conceptual information regarding Fog architectures and for a specific use case scenario. The real time implementation and performance evaluation to ensure the effectiveness of the architecture proposed, is reserved for future work.

\textbf{Conclusion:} The works presented in this section are mainly architectures only, focusing on scalability of IoT networks and reduction in processing/communication overhead. Implementation and evaluation are two key elements missing from these solutions. Similarly, coupling of SDN and synchronization of different VFs with orchestrator and control layer could lead to improvement in deployment of VNFs in IoT.

\subsection{Management of IoT using NFV}
This sub-section reviews management specific literature for function virtualization in IoT environments. Some of the solutions uses SDN technology besides NFV.
 
Batalle \textit{et al.} \cite{batalle2013implementation} integrates NFV and SDN to reduce cost in IoT, where centralized controller is responsible for routing which has a global view of the network. This work presents a novel design of a virtualized routing protocol using NFV infrastructure. It simply manages and reduces signaling overhead, particularly when inter-domain routing is required. The NFV implementation for virtualization of the routing function is done over an OpenFlow network. It aims to also reduce the number of connected and deployed devices, hence will reduce the cost as well. Just like OpenFlow, packet is inspected and if required, it is sent to the Floodlight controller which then takes decision after inspecting whether packet belongs to IPv4 or IPv6. Proposed solution is implemented using GEANT \cite{GEANT}, that offers infrastructure to emulate OpenFlow-based SDN solutions. As the amount of communication increases, the proposed solution is able to reduce the number of flow entries by 50\%, which improves performance and scalability. But to improve the robustness of the virtualized function, more evaluation are expected. The experiment leads to a number of open research questions, starting from implementation of dynamic routing protocols in the virtualized host, to different routing policy optimization.\par
Maksymyuk \textit{et al.} \cite{maksymyuk2017iot} adopts IoT-based network monitoring framework to manage the performance of 5G heterogeneous networks under different conditions. In this architecture, Radio Access Network functionalities are virtualized using NFV to simplify load balancing and spectrum allocation. On the other hand, the centralized intelligence of SDN controller is used to implement interference aware spectrum allocation. This allows better load balancing of smaller cells and manages user's mobility. This proposed framework has two main advantages. Firstly, only relevant data will be subscribed by each network operator that can improve the existing monitoring system. It also supports multiple Mobile Network Operators (MNOs). Secondly, the small size of transmittable data block generates less traffic overhead.
 
Zhang \textit{et al.} \cite{Zhang:2016:OPH:2940147.2940155} proposes an extension to OpenNetVM using Network Function (NF) management module that manages on-demand NFs in lightweight Docker containers. This is to facilitate various service providers, leveraging startup duration and memory consumption of CPUs. OpenNetVM supports flexible and high performance NFV architecture for a smart IoT platform, enabling increased interoperability among NFs. NF management module is an efficient and scalable packet processing architecture that enables dynamic manipulation of packets using service chains. The simulation result shows significantly high rate of throughput for packet transmission leveraging Data Development Kit (DPDK) \cite{DPDK} to improve performance I/O. This creates scope to render complex software based services for deep analysis within the network and data centers. This work may also remove the limitation of managing large volumes of IoT devices to some extent. However, on-demand NF deployment is limited to CPU cores.

\textbf{Conclusion:} In all the efforts mentioned in this subsection, third party services are involved to manage and facilitate the network topology. Usage of SDN controllers may also include management services from vendor specific organizations. Research community may work on developing SDN/NFV-based advanced real-time applications to manage and orchestrate IoT nodes in the context of knowledge-based 5G mobile networks.

\subsection{NFV-based Security solutions for IoT}
This sub-section presents different security solutions, which use NFV to implement security in IoT.
 
Massonet \textit{et al.} \cite{8114476} proposes an extended federated cloud networking architecture for edge networks and connected IoT device security. The security solution utilizes lightweight virtual functions and Service Function Chaining (SFC). The IoT gateways in the edge networks are responsible for implementing global security policy, by creating a chain of VFs for different purposes, such as, firewall and intrusion detection. They monitor the IoT devices for vulnerabilities and attacks, and isolate the device if it is detected. 
SFC is also responsible for flow management within the IoT network and with the federated cloud, which requires the cloud and IoT platforms to have appropriate infrastructure to support it. This is achieved by implementing a \emph{federation agent} at IoT controller or gateway level. The communication itself is done using REST API. 
The federated network manager sends configuration information to IoT network Controller, which is then forwarded to gateways for implementation.  Finally, the network controller exchanges information with the IoT proxy, which helps manage the data plane using OpenFlow protocol. To secure the IoT-Cloud network slices, a module is implemented inside the IoT network controller. Future work may incorporate enhancing scalability among IoT devices, and algorithms to preserve strong security \& privacy in the edge IoT network.

Al-Shaboti \textit{et al.} \cite{8432333} proposes novel IPv4 address resolution protocol (ARP) server providing NFV security service to defend against ARP spoofing attack, and network scanning. The work also proposes an SDN-based architecture for enforcing network static and dynamic access control of smart home IoT. All ARP requests pass through a virtualized trusted entity called ARP server. It is able to secure all ARP operations, eliminating the ARP broadcast messages, and easily legitimates ARP spoofing through ARP proxy by configuring the ARP server. The work resolves packet processing delay problem using high-speed packet processing technology. Such technologies include deep packet inspection (DPI), multi-core processor, carrier-grade operating system with Linux, and virtualization enabling the sharing of cores between applications. Only NFV-IoT related contribution is focused here. The design architecture includes local components like data plane,  NFV dispatcher, and local security services. Security agent, as one of remote components, takes input from user control plane, IoT policy manager, security services, and configures the SDN (Ryu) controller to enforce the corresponding network access control rules. NFV dispatcher receives all mirrored packets relaying from mirror port. Then forwards them to the corresponding security service based on the dispatcher list. Security agents extracts related information to direct security services for each flow. Based on the examination, security decisions/alerts are generated. IPv4 ARP server validation shows that it can protect ARP spoofing, and corresponding data plane deployment kit (DPDK) implementation performs well for the smart home IoT network. Future work will extend incorporating intrusion detection and prevention system into this architecture, and include IPv6 as a key enabler for IoTs.\par

\textbf{Conclusion:} NFV or SDN domains have different elements, applications, orchestration managers, virtual functions, communication APIs, etc. A malicious or compromised element in any of them may have serious effects on the whole system. For example, a malicious VNF by a compromised software vendor, a compromised hypervisor, or MANO component, could harm the entire network domain. If these elements are well secured than integrity, confidentiality, availability, access control, and accountability can be well preserved. Research work on access control \& packet inspection mechanisms, needs further investigation, specially for resource constrained IoT devices.

\section{Software-Defined Internet of Things (SDIoT)}

In this work we classify SDN-based IoT solutions and SDIoT solutions as two separate categories, with different scopes. SDIoT extends the Software-Defined (SD) approach to collect and aggregate data from network devices, sensor platforms, and cloud platforms. It uses sensing applications to provide standard API services for data acquisition, transmission, and processing. The SDN technology provides packet flow configuration for network devices enhancing network connectivity, hence SDN-based IoT is limited to network layer virtualization.  NFV implementation extends the network connectivity and security. The basic idea is to virtualize key NFs, and place them on commodity servers. Next step is to connect them via a flexible SD infrastructure managed through a unified orchestration system. For optimization, service provisioning, scalability, performance enhancement, and rapid deployment, the whole IoT ecosystem can be virtualized by SD paradigm. Hence, SDIoT solutions are not limited for a specific layer, but ranges from device up to application.\par
\begin{figure}[!t]
	\centering
	\includegraphics[width=\linewidth]{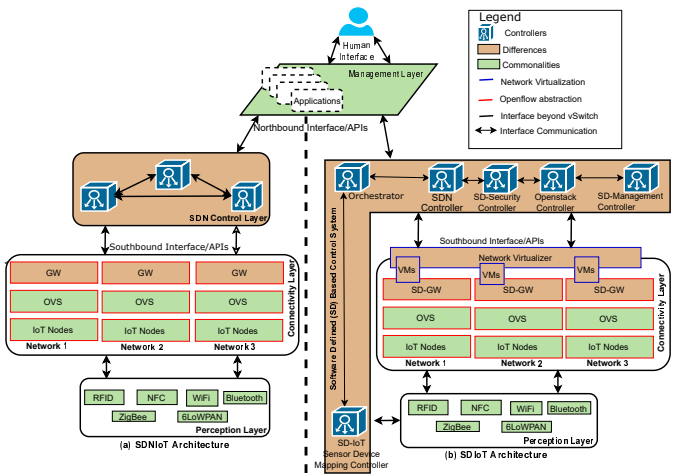}
	\caption{Difference between SDN-based IoT \& SDIoT architectures. Left figure shows the generic SDN based IoT framework, while right side shows a complete software define IoT design.}
	\label{fig:SDN_vs_SD}
\end{figure}

 
The difference between SDN-based IoT architecture (Figure~\ref{fig:SDN_vs_SD}a) is very subtle but significant as compared to those of SDIoT (Figure~\ref{fig:SDN_vs_SD}b). Control layer is improved by customizing more domain-specific SD-controllers, each executing specific tasks within SDIoT architecture. This reduces the burden on single controller. The control layer is extended not only horizontally, but also vertically. Hence, the function virtualization orchestrator becomes an integrated part of control layer. Protocols/APIs for SDIoT framework varies upon the nature of communication, and the type of IoT devices connected to it. A widely used OpenFlow protocol already exists to communicate between SD-controller and OF-switch, but it needs to extend it's capabilities to communicate with IoT devices beyond virtual switches. Application and management layer communicates with the connectivity layer through the NBI. This layer can also have a management specific framework, which can enforce different policies through the programmable interface for SD-controllers to execute. This framework can also enable different virtual functions at different layers of SDIoT network for groups of different nodes. SD-controllers enforce different policies. Enforcement of these policies is pushed through virtual functions from multiple controllers. For example, SD-Security and SD-Management controller enforce security policies and management related policies, respectively. OpenStack controller orchestrates network slicing. SDN controller configures OF-switch to install data flows, best routing paths, and network control. More controllers can be assigned based on the nature of network management objectives and network performance. Orchestrator is responsible for configuring different SD-controllers on-demand, not only along the horizontal control plane but also vertically. SDIoT controller enforces IoT device specific management rules. It works in collaboration with the orchestrator and other SD-controllers to enhance the communication of perception layer with the control layer via connectivity layer. It eventually enables seamless end-to-end IoT device communication in an SDIoT environment.

The following sub-sections present different architectural, management, and security solutions exploiting different SD-controllers. Table~\ref{tab:TableVI} summarizes \& categorizes SDIoT architecture, management, and security related literature.

\subsection{Architecture Solutions}
This section discusses SDIoT architectural solutions, using multiple controllers for providing different services. Remote configuration of networks and efficient data retrieval has been one of the core challenges of big data analytics for smart cities. Few efforts have been done to use SD and IoT potential to counter these issues.

Din \textit{et al.} \cite{8110222} proposes an SDIoT architecture, which consists of data collection \& management controller. The data passes through Data Processing Layer, Data Management Layer, and Application Layer. The architecture uses multiple SD-controllers/SD-gateways. Data is collected through a novel data collection algorithm, from various IoT-enabled embedded devices. The aggregated data, via various Aggregator Points (i.e. Zone, Local, and Global), is passed on to Data Processing \& Management layers, for real-time data processing and extraction. Since IoT devices generates large volumes of data, the proposed system utilizes Hadoop Distributed File System for data storage \& manipulation purpose. The work contributes by inserting a novel data processing algorithm setting threshold limit values for every data set. The work also uses Information Centric Network \cite{7986623} and Named Data Network \cite{8308426} potentials to fulfill its requirements. The simulation result shows promising aspects. HDFS works significantly well analyzing data with high throughput and less processing time, even though the throughput and processing time may still be improved using cluster based Hadoop system with efficient scheduling mechanisms.

Liu \textit{et al.} \cite{liu2015software} proposes an SDIoT architecture to separate smart urban sensing applications from the existing physical infrastructure, because most of the underlying network element (e.g. sensor nodes) are not SDN-enabled. The control logic of these devices is encapsulated in hardware. The authors divide the entire framework into three layers, i.e. physical infrastructure layer (sensors, smart phones,gateways,etc.), control layer (SD controllers), and application layer (IoT applications). SD controllers are used to manage specific configuration for each hardware resource and provide interface to standard API services for data manipulation. Each of these controllers can be replicated to enhance its robustness and can be physically placed anywhere for resource usage optimization. Basically, data is first aggregated in the sensor platform and passed on to the network infrastructure to calculate the best routing path for end-to-end IoT device data transmission, using SDN-enabled networks. Every sensor platform in SDIoT architecture is facilitated with more than one sensors of similar or different types and shared by many applications. Sensor controller has the global view of the underlying physical infrastructure and capable of activating/deactivating sensors dynamically.The forwarding devices are OpenFlow-enabled and programmable, and the SDN controllers are responsible for scheduling packet flow tables for forwarding devices, and smart traffic steering. Hence, optimizing network resource usage. On the other hand, cloud platform allows urban sensing data to be stored and processed. Cloud controller monitors and maps the underlying server resource pools. Although the architectural design is supported with case studies and qualitative investigations only, it shows promising possibilities to improve network resource utilization as well as dynamic data optimization, processing, and transmission. Future work may focus on controller inter-communication and resource utilization.\par 
By applying the fundamental SD features like centralization, virtualization, optimization, another similar approach is taken by Xu \textit{et al.} \cite{xu2016toward}. They proposed an IoT-based software defined Smart Home (SDSH). It supports openness, virtualization, and centralization, integrating the heterogeneous network devices in smart home domain. The entire platform has been divided into three main layers namely controller layer, intelligent hardware layer, and external service layer. The controller acts as a management layer providing compatibility and API support to different smart devices and third party services, respectively. The APIs through the SDN controllers from the control layer handle the communication and interaction between the peripheral IoT devices. These APIs are also responsible for IoT device registration based on their specifications. The architecture also uses virtualization technology to maintain uniform virtual abstraction of hardware computing, storage, and network resources of the whole smart home ecosystem. In addition, it uses virtual network function for access control mechanisms, firewall, load balancing, etc. The literature only reviews key technologies and challenges of SDSH. Although the overall architecture shows promising aspects, simulation or real-time experiment should be carried out in the future to prove effectiveness of the solution.\par
Hu \textit{et al.} \cite{7324414} proposes a dynamic controllable solution for Software Defined Industrial IoT (SDIIoT) with SDN features in it. The solution emphasizes on application specific holistic performance approach of network nodes like field devices, gateways, and sensor cloud in respect to connectivity and interoperability. The proposed architecture has three different network building blocks: IIoT sensor cloud, IIoT gateway, and IIoT field device. The control plane is responsible for configuring these network nodes, and uses different controllers for it. QoS controller enforces QoS policies for the network backbone and field WSN. Network controller handles topology management and data updates. Timeliness is dealt by the data synchronization controller, and security controller enforces security schemes for these network nodes. An additional Data Manger module provides data management services, and control module implements control plane functions. The authors uses Floodlight controller for configuring open virtual switch via IIoT gateway for best routing paths during real-time data transmission, then compare their results with Amazon AWS and sensor cloud server. They show that latency can be reduced by 30\% to 38\%. Moreover, system is reliable and success rate is 100\% because of QoS mechanism in CoAP protocols. Future work can explore and exploit the SDIIoT from big data perspective employing problem specific networking techniques.\par
Wan \textit{et al.} \cite{7467436} proposes a Software-Defined Industrial Internet of Things (SDIIoT) architecture utilizing SDN technology and industrial cloud. The architecture consists of three layers. Physical layer consists of various kinds of hardware devices such as sensor, gateway, switch, router, etc. Control layer manages the physical infrastructure underlying it. It includes SD-controller and SDN controller for processing specific tasks and configure the switches/gateways within its network via SBI. NBI allows applications from the application layer to implement decisions on SD-Controllers based on industry enterprise needs. Application layer also provides different kinds of APIs to monitor equipment fault and usage, and product processing. The proposed SDIIoT architecture also provide three major services: data collection, data transmission, and data processing. Data collection is processed through the application layer APIs, where data sensed is transmitted either via wireless or wired networks. Data processing occurs simultaneously to process one or multiple IoT devices. Decision making is autonomous while the data processing is software defined.
As the system would deal with large scale big data, the SDIIoT service mechanisms require high-quality data process mechanisms/algorithms, which the authors aim to develop in future. The authors also provides security suggestions related to illegal access, vulnerabilities caused due to IoT device mobility and large number of sensor/IoT nodes, which are potential directions for research community.

\textbf{Conclusion:} The contributions in this sub-section include IoT/IIoT concepts with SD features. The solutions mainly focused on incorporating APIs in the application layer to enforce decision rules on SD-Controllers and to exploit network virtualization features, eventually providing global view of IoT nodes beyond virtual switches. Future work may focus on integration of VFs specialized for different controllers and their distributed placement in the network. Moreover, the distribution of different controllers in the networks may improve performance and reduce the communication latency with IoT devices. In this regard, inter controller communication may also require further improvement and standardization.

\begin{table*}[]
	\centering
	\caption{Software Defined IoT Solutions and their Classification.}
	\label{tab:TableVI}
	\setlength\tabcolsep{2.5pt}
	\begin{tabularx}{\linewidth}{|>{\hsize=0.6\hsize}Y|>{\hsize=0.7\hsize}Y|>{\hsize=1.2\hsize}Y|>{\hsize=0.7\hsize}Z|>{\hsize=1.5\hsize}Y|>{\hsize=1.3\hsize}Y|}\hline 
		
		\textbf{Literature \& Classification} & \textbf{Objective(s)} & \textbf{Solution}  & \textbf{* Control Plane Architecture}	& \textbf{Benefit(s)} & \textbf{Limitation(s) / Future Work} \\\hline
		
		Din et~al. \cite{8110222} \newline[4pt]\textit{Architecture} &
		Data sensing, collection, \& processing.\newline[4pt]
		Scalability \& availability.	&
		SDIoT architecture to analyze data of smart cities.	&
		Distributed	&
		Uses a Hadoop ecosystem for load balancing.\newline[4pt]
		Data collection done using SDN and NDN. &
		Complex scheduling algos needed for cluster based Hadoop systems.\\\hline
		
		Liu et~al. \cite{liu2015software} \newline[4pt]\textit{Architecture}	&
		Sensing \& robustness	&
		SDIoT architecture for smart urban sensing.	&
		Distributed	&
		Dynamic data optimization, processing, and transmission.	&
		Multiple application configuration persists on shared sensor platform.\\\hline
		
		Xu et~al. \cite{xu2016toward}  \newline[4pt]\textit{Architecture} &
		Scalability, Mobility, Openness.	&
		Smart Home IoT device integration with with SDN-based services.	&
		Centralized	&
		Virtualization to simplify heterogeneity \& complexity of diff. SDSH protocols.	&
		Architectural design only. No implementation or evaluation.\\\hline
		
		Hu et~al. \cite{7324414}  \newline[4pt]\textit{Architecture}	&
		Reliability, scalability, security, \& QoS.	&
		SD-IIoT architecture to manage data exchange and delay.	&
		Distributed\newline[4pt] \textit{(FloodLight)}	&
		Application specific approach for node performance, connectivity, \& interoperability.\newline[4pt]
		Focus on network controllability: processing, queuing, transmission, and delays.	&
		Optimization for more than 10 parallel connections not possible.\\\hline
		
		Wan et~al. \cite{7467436}  \newline[4pt]\textit{Architecture}	&
		Reliability, standardization, \& security.	&
		SD-IIoT architecture for seamless data processing.	&
		Distributed	&
		SD-data collection,transmission, \& processing mechanisms.\newline[4pt]
		Provides solution for: illegal access, and vulnerabilities caused by IoT device mobility, \& large crowds of IoT nodes.	&
		Limited evaluation of the proposed solution.\\\hline
		
		Nastic et~al. \cite{6984208}  \newline[4pt]\textit{Management}	&
		Configuration, operation, and access control of cloud system.	&
		Fleet management system using SDIoT cloud.	&
		Distributed	&
		Overall resource usage optimization.\newline[4pt]
		Elastic policy based configuration.\newline[4pt]
		Cost awareness.	&
		Limited implementation \& evaluation.\newline[4pt]
		Research on run-time SDIoT governance, \& edge network resource usage required.\\\hline
		
		Kathiravelu et~al. \cite{Kathiravelu:2015:CMP:2836127.2836132}  \newline[4pt]\textit{Management}	&
		Sensing, security, \& scalability.	&
		A middleware solution for context-aware smart buildings using SD WSN.	&
		Centralized	&
		Avoids single point of failure.\newline[4pt]
		Fast response to dynamic changes.	&
		Prototype is limited to single building.\\\hline
		
		Wu et~al. \cite{wu2015ubiflow}  \newline[4pt]\textit{Management}	&
		Scalability \& reliability.\newline[4pt]
		Mobility.	&
		Distributed overlay structure to support mobility management, and dynamic flow control.	&
		Distributed\newline[4pt]\textit{(FloodLight)}	&
		Mobility management, Handover optimization, and Distributed control.	&
		Flow-scheduling optimization issues concerning backbone network.\\\hline
		
		Jararweh et~al. \cite{jararweh2015sdiot}  \newline[4pt]\textit{Management} & Scalability, Heterogeneity, Agile, \& Inexpensive.	&
		SD solution for IoT to forward, store, \& secure data.	&
		Distributed\newline[4pt]\textit{(Multiple SDN controllers)}	&
		Multiple SD application modules to facilitate IoT network.	&
		Architectural design only. No implementation or evaluation.\\\hline
		
		Salman et~al. \cite{SDIoTSecurity_Iqbal}  \newline[4pt]\textit{Security} &
		Security, Privacy, \& Connectivity.	&
		Security solution for SDIoT utilizing SDN \& NFV technologies.	&
		Distributed	&
		Slicing techniques.\newline[4pt]
		Cloud based edge computing.\newline[4pt]
		Low latency, high throughput \& scalability with location awareness.	&
		Inter-access controller connectivity challenges.\\\hline
		
		Darabesh et~al. \cite{SDSecurityExpFramework_Iqbal}  \newline[4pt]\textit{Security}	&
		Enhanced security and reduced cost of security cost operations.	&
		SD-security solution.	&
		Centralized	&
		Virtualized SD-security elements: host, switch, and controller.\newline[4pt]
		Context-aware security solution.\newline[4pt]
		Supports security component configuration.	&
		Traffic overhead optimization challenges.\\\hline
		
		\multicolumn{6}{>{\hsize=\dimexpr6\hsize+6\tabcolsep+\arrayrulewidth\relax}X}{* Literature either does not mention any controller or assumes generic controller.}\\
	\end{tabularx}
\end{table*}
	
\subsection{Management Solutions}
Managing and configuring a diverse range of IoT devices can be a challenging task. In order to reap benefits of network programmability and efficient resource utilizations a few works have focused on SD-IoT management solutions.

The work done by Nastic \textit{et al.} \cite{6984208} applies SD in IoT, where they try to abstract the IoT resources in cloud by encapsulating them in software defined APIs. The proposed system directly interacts with the underlying physical IoT infrastructure. The main component in the system is the SD-gateway which implements predefined algorithms specified for tracking vehicles utilizing cloud. The objectives are to provision configuration, access, and operation of IoT cloud systems for a unified view. The authors use a vehicle fleet management system as a usecase. The architecture presents fundamental building blocks of SDIoT cloud systems by automating provisioning processes and supporting configuration models, eventually trying to make simple and flexible customization for IoT cloud for operation managers. On the other hand, exchange of raw IoT data in cloud needs a lot of computational resources and bandwidth. The future plan is to consider techniques and mechanisms to support runtime governance of SDIoT systems, enable SDIoT to optimize resource usage of edge networking, and allowing policy based automation of security and data-quality of SDIoT systems.\par 
Kathiravelu \textit{et al.} \cite{Kathiravelu:2015:CMP:2836127.2836132} proposes an architecture for Software Defined Building (SDB) \cite{6224230}, using smart clusters. This enables communication among IoT appliances within a multi-building campus. SDB is a platform to enhance the programmability and re-usability of IoT appliances. It also uses Software Defined Sensor Network \cite{6324377} to manage  communication mechanisms between sensors \& IoT appliances, and system policy implementation. The prototype CASSOWARY is partially Software Defined because it works on top of traditional SDN environment. It has a two layer architecture. Network layer has control and data plane, whereas, appliance layer manages the integration of smart appliances. The addition of IoT device SD-Controllers to the SDN controller, allows fast response to dynamic changes. Sensors and IoT nodes are connected to the SDN-enabled switches in the data plane. Different controllers  deployed are physically distributed in a cluster, which avoids a single point of failure. The message broker in the control plane assists SDN controller to distribute flow information and orchestrate the smart appliances and sensors.
A full scale deployment over real world scenario is complex and authors have left it for future. This may also include energy efficient and access control mechanisms for different smart devices.

Instead of using completely centralized controllers in the IoT based urban mobile networks, Wu \textit{et al.} \cite{wu2015ubiflow} introduces a distributed overlay structure to support ubiquitous mobility management and dynamic flow control where the entire SDIoT network topology is divided into different geographic chunks or clusters. Using a distributed hashing algorithm, each controller is assigned to a single IoT platform to solve the scalability problem. The authors focus on logical centralization of controllers while they are physically placed at different locations. An orchestration controller is used to communicate with local controllers. All controllers are OpenFlow compatible, coordinating with the mobility management of each mobile sensor platform. As the mobile sensor platform finds the gateway managed by one of new local controllers on the move, it sends the event details to the orchestration controller. It then coordinates with the initial/original controller and the local new controller, to provide smooth handover mechanism. The authors emphasize that this process of mobility management supports SDIoT paradigm. Moreover, a unique distributed protocol is used among these controllers independently to handle the single point of failure. However, for backbone network, further provisioning of flow-scheduling optimization is considered as future work. 

To address the needs of heterogeneous nature of IoT applications and objects, Jararweh \textit{et al.} \cite{jararweh2015sdiot} proposes a comprehensive SDIoT framework, to provide an improvement over IoT management layer. This model enhances several important aspects like security, storage, and traffic forwarding. It has three main components. First, physical layer deals with all physical devices like sensors, servers, switches/routers and security hardware. Second, control layer is the core of the proposed prototype to manage and coordinate among different SD-controllers, i.e. IoT controllers, SDN controllers, SDStore controllers, and SDSec controllers to abstract the management and control operations from the underlying physical infrastructure. Each of these SD-controllers run specific tasks in the control layer. Third, application layer through NBIs combines fine-grained user applications to facilitate access control and data storage mechanisms. The administrator is able to configure them through the Eastbound Interface and the inter-controller communication may occur using the Westbound APIs. Additional controllers can be added to tackle sophisticated load balancing and inconsistency issues and to deliver fast response time for many requests within the network. The authors in this prototype only exploit the ideas of SDN, SDStore, and SDSec to build the architecture but these SD-controllers' detail functional elements is planned to be developed in the future.

\textbf{Conclusion:} Most of the research contributions mainly focus on extending APIs in the application layer to enforce decision rules on SD-Controllers, SD-gateways, and to exploit network virtualization features. However, in presence of multi-vendor solutions at application, control, and data layers, standardization for communication interfaces (NBI) becomes very critical. Moreover, functionalities specific to IoT devices should also be part of overall management architecture such as mobility and resource management, etc.

\subsection{Security Solutions} 
Enforcing policies for security and access control in large scale networks, can be made easier by programmable interfaces. An efficient solution can be defined by adding a dedicated security controller in the SD infrastructure for IoT network. Below we discuss solutions, which have focused on a similar concept.
  
Salman \textit{et al.} \cite{SDIoTSecurity_Iqbal} discusses the IoT requirements in terms of security and privacy. In addition, an IoT Software Defined security framework is proposed where software defined and virtual function technologies are used for virtualization, and further slicing techniques are used for isolation of the network, blended along with Mobile Edge Computing (MEC). They provide cloud based edge computing services at the user’s proximity, which gives valuable benefits such as location awareness, low latency, augmented reality, high throughput \& scalability, etc. The architecture consists of six layers. There are two types of controllers. Core controller acts as a global network OS, while access controller provides a dynamic control model to support IoT device communication. They are located in the core network and access network, respectively. The devices in the data plane are connected with access points. Each IoT device after registration is provided with initial credentials, and assigned a security level depending on it's capabilities (computation, storage, energy), which affects its authentication. The scheme has Authentication and Authorization Delegation requests passing through different layers based upon IoT device type. IoT devices are also tested using the Automated Validation of Internet Security Protocols and Applications (AVISPA) tool, which uses High Level Protocols Specification Language (HLPSL), a high-level role-based language for security protocol description. Authors have evaluated against only few back-end attack modules to test the security goals. As the overall system is very complex, the evaluation is limited. Future work on inter-controller connectivity and seamless integration of modules may enhance the system.

Darabseh \textit{et al.} \cite{SDSecurityExpFramework_Iqbal} addresses the challenges of providing multiple levels of protection and efficiency in an SD environment. They propose a centralized yet flexible security solution by abstracting the security mechanisms from the hardware to a software layer, providing virtualized testbed environment for Software Defined Security (SDSec) systems, grouped under Software Defined System (SDSys). The system is software defined because the SDSec\_Controller (i.e. software-based POX controller) has the ability to manage diverse data place resources regardless of their vendors. The framework uses Mininet simulator to create a virtual environment for emulating different forms of SDSec policies and test their performance under different scenarios.
The core customized components are SDSec\_Host, SDSec\_Switch, and SDSec\_Controller. The framework uses Catbird \cite{catbird} for policy deployment on hardware assets. It is completely software-based, and unrestricted to hardware, easy to scale, and can adapt it to new changes. It is able to create on-demand new VMs, without the need for manual intervention. Hence, SDSec imports Catbird into its security framework to virtualize the security functions, and to increase the discoverability of the problems and abnormal actions \& activities. 
The authors aim to extend the SDSecurity by developing a distributed controller which may reduce the overhead and enhance performance. More security controls inside each SD controller can be further added in future.

\textbf{Conclusion:} The solution presented here mainly focused on preventing DDoS attacks, access \& congestion control mechanisms, and slicing techniques. Future work may include developing APIs that may interact simultaneously through the NBI, SBI, and E/WBI, in order to enforce security decisions \& rules to the SD controllers/gateways. It may also be able to exploit network virtualization features to assign VNFs for preventing different threat vectors. The solutions should be able to defend against a wide range of threats, providing global view of IoT nodes beyond virtual switches.

\section{Future Research Directions}
The fundamental objective of this article is to collect, categorize, and analyze different software defined and virtual function solutions for Internet of Things. From this analysis, we have identified key challenges and possible research directions in this domain. The summaries of these are given at the end of each sub-section in the paper, however, in this section, we elaborate them in greater depth. The success of SDIoT requires improvement in different layers of the overall system, hence we classify them accordingly.

\textbf{Application Layer:} This is the top most layer, and mainly responsible for user/administrator interaction, and other generic application models for enforcing and configuring different policies in lower structure. Some of the core research directions are as follows: 

\begin{itemize}
	\item In past SDN applications have been written specific to certain functionalities and only for specific controllers. This creates a major bottleneck as there is no standardized application development framework available. Such a framework will be highly beneficial for both research community to build test applications and also for industry in rapid deployment of SD-IoT networks.
	
	\item Similarly existing controllers mostly use REST API for communication with application layer. Unlike SBI there has been little to no effort in standardizing NBI. This effort will certainly be an important step towards widespread adoption and development. An important research element in this regard is to allow diversified application and controller capabilities. As applications and controllers are both specific to different functionalities in SD-IoT framework, hence the standardized NBI must be flexible enough to accommodate different types of communications.
	
	\item Most of the focus with regards to security has been on flow security and attacks on networks. Hence security policy enforcement has been extensively studied. However, security of application modules is also as important as the prior. A compromised application module can mis-configure and severely compromise an whole SD-IoT ecosystem.	
\end{itemize}

\textbf{Control Layer:} This is the main focus area of SD-IoT and will require major research and contribution efforts. It is not possible to list all potential directions, hence, we list the major concerns for control layer in SD-IoT. 

\begin{itemize}
	\item Communication among different elements of control layer is an important aspect. In traditional software defined network, the controllers independently controlled a domain, and those in hierarchy would use proprietary methods of inter-controller communication. However, in SD-IoT there are multiple types of controllers for some domain, which rely on each other for complete working. In addition, the control layer may use multi-vendor solutions, hence a standardized interface is an important research direction. Communication with other domain’s controller may also be investigated for efficient \& optimized communication. \cite{yin2012} did an interesting work in this regard, which may be a starting point.
	\item In traditional SDN, single point of failure of control layer is avoided by back-up controller. However, due to diversity in SD-IoT controllers, having a back-up controller of each controller may require investigation for deployment costs and complexity. This also impacts scalability, hence more novel architecture for controller redundancy may provide better solutions.
	\item SBI is a major element of control layer. OF has been a defacto interface for network controller to data plane communication. In light of diverse SD controllers, suitability of OF may need reevaluation. SBI which can effectively work for all types of controllers and devices will be another interesting direction. At the same time, the SBI should be able to reach IoT devices beyond vSwitches. OF does not connect hosts, but only allows flow installation on switches. In an IoT networks the mobile devices and AP may also need configuration and other policy enforcements. This requires enhancements to OF or new SBIs which can reach beyond vSwitches.
	\item Efficient use of network function virtualization is also a key research direction. Function chaining for various controller processes may enhance the performance, and allow better control in network. As the vertical control layer implements VNFs, their orchestration with the horizontal controllers is also an open research challenge.
	\item In addition to other control layer challenges, security of control layers itself, its elements, and communication is extremely important. The security controllers should not only focus on security of data plane and network devices, but should also ensure logical element security. Research in this direction will have a major impact on SD-IoT networks.
\end{itemize}

\textbf{Controller perspectives:} Controller Perspective: SD-IoT will consist of a number of controllers designed for specific operations. This will allow a number of research directions to be explored.
\begin{itemize}
	\item Placement of a controller or other control layer elements is a less researched area, mainly due to a single network controller. In SD-IoT networks, the number of controllers and topological structure of IoT devices may require a more close look at the placement in topology for different controllers.
	
	\item IoT networks will compromise of hundreds of devices (if not thousands) in a single SD domain. Hence the scalability of controllers is an important factor. This will include not only scalable architectures, but also languages, thereby capabilities, storage, and processing at controllers. As there are multiple types of controllers, hence, scalability and coupling at a large scale will be very interesting research direction.
	
	\item Synchronization of controllers and their policies will also be an interesting challenge. Furthermore, it will be equally interesting to evaluate the requirements of domain and then utilize only those types of SD controllers which are required. Vertical versus horizontal deployment of controllers and associated VNFs may also present interesting design options.
	
	\item Controller virtualization is an important element in software defined IoT systems. Virtualizing multiple controllers and coordination among them is a challenging task. Similarly, placement of virtualized elements in core or edge network will be an interesting research issue.	 
\end{itemize}

\textbf{Management Perspective:} The nature and properties of IoT networks has highlighted some newer research challenges, which were not evident in traditional SDNs. In a complete SD-IoT system, these will require significant attention from research community.

\begin{itemize}
	\item \emph{Mobility:} In a single SD-IoT domain there may be multiple edge networks, with dozens of diverse mobile IoT devices with high mobility and limited resources. Some solutions have tried to address mobility in SDNs, however, in a hybrid adhoc-infrastructure environment with different physical layer technologies, it will present new research dimensions. Efficient and quick topology discovery in mobile domain, path configuration, hand-over and other scalability challenges should be further investigated.
	
	\item \emph{Device configuration:} The edge and access network in an SD-IoT network will comprise of heterogeneous mobile devices. A major challenge is to configure them according to policy dictated by the application layer. This also requires significant research before a unified framework can be developed.
	
	\item \emph{Virtual functions:} Virtualization of different network functions will be an integral part of SD-IoT ecosystem. Hence, their management in control lance, distribution, virtualization, and integration with other layers \& APIs is a major research area.
\end{itemize}

\textbf{Technology Interaction \& Complexity: } Most of the previous challenges \& directions also deal with complexity of overall architecture, but the research community needs to look at integration of other technologies in the overall ecosystem, such as fog/edge computing, cloud computing, crowd sensing, Blockchain, etc.

\begin{itemize}
	\item Crowd sourcing techniques can benefit extensively from SDIoT networks. The functions for task advertisement, auction, bidding, and offloading can be easily implemented through virtual functions, and orchestrated by a crowd sourcing controller placed at the edge node. Such an architecture, can enable rapid deployment of sourcing tasks and collection of data. However, this will certainly require further research in the specific controller design, virtualization of such controllers, and security among other challenges. This will also increase the complexity of overall SDIoT frame work, hence requiring more scalable systems.
	\item Blockchain is a relatively new area for IoT, but may prove to be extremely beneficial in financial transactions and other private Blockchain trades. Potential research directions may involve virtualization of complete peers/mines, offloading of complex mathematical functions \& proof of work to other nodes via virtual functions, virtualization of Blockchain ledger, etc. SDIoT may also pave the way for hybrid Blockchains for Internet of Things. 
\end{itemize}

\section{Conclusion}
Software defined networks have seen extensive deployment in data centers and core networks, where they have been mostly used for flow optimization and related policies. The recent advancement in Internet of Things has created a keen interest of research community as well as industry to integrate SDN in IoT networks. Similarly, the virtualization in terms of networks, functions, and devices has also seen significant contributions in recent past. In this article, we have reviewed both SDN and virtualization techniques for IoT, and classified them into different types of solutions. SDN is limited to virtualizing the network layer of the stack where the IoT network traffic flow is optimized. Mostly the solutions aim at providing SDN services to resources constrained devices, provide configuration services, or address security threats. Some works have involved function virtualization to implement common network functions in logical domain. An important factor to note is the future of IoT will not only be limited to SDN and isolated virtual functions. The later part of the paper emphasizes on software defined IoT, which is a comprehensive solution, by incorporating controllers for different purposes in the control layer. This also integrates orchestration of virtual functions, as part of the vertical control layer. Additionally, we have presented a number of future research directs in this regard.

\bibliographystyle{IEEEtran}
\bibliography{library}

\begin{thebibliography}{100}
\providecommand{\url}[1]{#1}
\csname url@samestyle\endcsname
\providecommand{\newblock}{\relax}
\providecommand{\bibinfo}[2]{#2}
\providecommand{\BIBentrySTDinterwordspacing}{\spaceskip=0pt\relax}
\providecommand{\BIBentryALTinterwordstretchfactor}{4}
\providecommand{\BIBentryALTinterwordspacing}{\spaceskip=\fontdimen2\font plus
\BIBentryALTinterwordstretchfactor\fontdimen3\font minus
  \fontdimen4\font\relax}
\providecommand{\BIBforeignlanguage}[2]{{%
\expandafter\ifx\csname l@#1\endcsname\relax
\typeout{** WARNING: IEEEtran.bst: No hyphenation pattern has been}%
\typeout{** loaded for the language `#1'. Using the pattern for}%
\typeout{** the default language instead.}%
\else
\language=\csname l@#1\endcsname
\fi
#2}}
\providecommand{\BIBdecl}{\relax}
\BIBdecl

\bibitem{IoT_Iqbal}
S.~Elbouanani, M.~A.~E. Kiram, and O.~Achbarou, ``Introduction to the internet
  of things security: Standardization and research challenges,'' in \emph{Proc.
  of Int. Conf. on Information Assurance and Security}, December 2015, pp.
  32--37.

\bibitem{bizanis2016sdn}
N.~Bizanis and F.~A. Kuipers, ``{SDN} and {V}irtualization solutions for the
  {I}nternet of {T}hings: A {S}urvey,'' \emph{IEEE Access}, vol.~4, pp.
  5591--5606, 2016.

\bibitem{7387427}
A.~Blenk, A.~Basta, J.~Zerwas, and W.~Kellerer, ``Pairing sdn with network
  virtualization: The network hypervisor placement problem,'' in \emph{Proc. of
  IEEE Conf. on Network Function Virtualization and Software Defined Network},
  November 2015, pp. 198--204.

\bibitem{8187644}
F.~Bannour, S.~Souihi, and A.~Mellouk, ``Distributed sdn control: Survey,
  taxonomy, and challenges,'' \emph{IEEE Commun. Surveys Tuts.}, vol.~20,
  no.~1, pp. 333--354, 2018.

\bibitem{7243304}
R.~Mijumbi, J.~Serrat, J.~L. Gorricho, N.~Bouten, F.~D. Turck, and R.~Boutaba,
  ``Network function virtualization: State-of-the-art and research
  challenges,'' \emph{IEEE Commun. Surveys Tuts.}, vol.~18, no.~1, pp.
  236--262, 2016.

\bibitem{7929710}
A.~L. Aliyu, P.~Bull, and A.~Abdallah, ``Performance implication and analysis
  of the openflow sdn protocol,'' in \emph{Proc. of Int. Conf. on Advanced
  Information Networking and Applications Workshops}, March 2017, pp. 391--396.

\bibitem{7073808}
N.~Omnes, M.~Bouillon, G.~Fromentoux, and O.~L. Grand, ``A programmable and
  virtualized network it infrastructure for the internet of things: How can nfv
  sdn help for facing the upcoming challenges.'' in \emph{Proc. of Int. Conf.
  on Intelligence in Next Generation Networks}, February 2015, pp. 64--69.

\bibitem{Schaffrath:2009:NVA:1592648.1592659}
G.~Schaffrath, C.~Werle, P.~Papadimitriou, A.~Feldmann, R.~Bless,
  A.~Greenhalgh, A.~Wundsam, M.~Kind, O.~Maennel, and L.~Mathy, ``Network
  virtualization architecture: Proposal and initial prototype,'' in \emph{Proc.
  of the ACM Workshop on Virtualized Infrastructure Systems and
  Architectures}.\hskip 1em plus 0.5em minus 0.4em\relax ACM, 2009, pp. 63--72.

\bibitem{5183468}
N.~M. M.~K. Chowdhury and R.~Boutaba, ``Network virtualization: state of the
  art and research challenges,'' \emph{IEEE Commun. Mag.}, vol.~47, no.~7, pp.
  20--26, 2009.

\bibitem{7350211}
Y.~Li and M.~Chen, ``Software-defined network function virtualization: A
  survey,'' \emph{IEEE Access}, vol.~3, pp. 2542--2553, 2015.

\bibitem{7218418}
L.~Galluccio, S.~Milardo, G.~Morabito, and S.~Palazzo, ``Sdn-wise: Design,
  prototyping and experimentation of a stateful sdn solution for wireless
  sensor networks,'' in \emph{Proc. of IEEE Int. Conf. on Comp.Comm.
  (INFOCOM)}, April 2015, pp. 513--521.

\bibitem{6385039}
S.~Costanzo, L.~Galluccio, G.~Morabito, and S.~Palazzo, ``Software defined
  wireless networks: Unbridling sdns,'' in \emph{Proc. of European Workshop on
  Software Defined Networking}, October 2012, pp. 1--6.

\bibitem{6324377}
T.~Luo, H.~P. Tan, and T.~Q.~S. Quek, ``Sensor openflow: Enabling
  software-defined wireless sensor networks,'' \emph{IEEE Commun. Lett.},
  vol.~16, no.~11, pp. 1896--1899, November 2012.

\bibitem{6838365}
Z.~Qin, G.~Denker, C.~Giannelli, P.~Bellavista, and N.~Venkatasubramanian, ``A
  software defined networking architecture for the internet-of-things,'' in
  \emph{Proc. of IEEE Network Operations and Management Symposium}, May 2014,
  pp. 1--9.

\bibitem{8169853}
S.~Bijwe, F.~Machida, S.~Ishida, and S.~Koizumi, ``End-to-end reliability
  assurance of service chain embedding for network function virtualization,''
  in \emph{Proc. of IEEE Conf. on Network Function Virtualization and Software
  Defined Networks}, November 2017, pp. 1--4.

\bibitem{8326277}
I.~Hwang and D.~Shin, ``Application level network virtualization using
  selective connection,'' in \emph{Proc. of Int. Conf. on Consumer Electronics
  (ICCE)}, January 2018, pp. 1--2.

\bibitem{8323847}
T.~Banchuen, K.~Kawila, and K.~Rojviboonchai, ``An sdn framework for video
  conference in inter-domain network,'' in \emph{Proc. of Int. Conf. on
  Advanced Communication Technology}, February 2018.

\bibitem{7496952}
Z.~Shu, J.~Wan, J.~Lin, S.~Wang, D.~Li, S.~Rho, and C.~Yang, ``Traffic
  engineering in software-defined networking: Measurement and management,''
  \emph{IEEE Access}, vol.~4, pp. 3246--3256, 2016.

\bibitem{Gude:2008:NTO:1384609.1384625}
N.~Gude, T.~Koponen, J.~Pettit, B.~Pfaff, M.~Casado, N.~McKeown, and
  S.~Shenker, ``Nox: Towards an operating system for networks,'' \emph{ACM
  Comput. Commun. Rev.}, vol.~38, no.~3, pp. 105--110, 2008.

\bibitem{8447647}
V.~Jara and Y.~Shayan, ``Latency measurement in an sdn network using a pox
  controller,'' in \emph{Proc. of IEEE Canadian Conf. on Electrical Computer
  Engineering (CCECE)}, May 2018, pp. 1--5.

\bibitem{6838330}
K.~Phemius, M.~Bouet, and J.~Leguay, ``Disco: Distributed multi-domain sdn
  controllers,'' in \emph{Proc. of IEEE Network Operations and Management
  Symposium}, May 2014, pp. 1--4.

\bibitem{ODL}
\BIBentryALTinterwordspacing
``{OpenDaylight: A Linux Foundation Collaborative Project},'' [Accessed
  15-January-2019]. [Online]. Available: \url{https://www.opendaylight.org}
\BIBentrySTDinterwordspacing

\bibitem{FL}
\BIBentryALTinterwordspacing
``{A Java based OpenFlow Controller},'' [Accessed 15-January-2019]. [Online].
  Available: \url{www.projectfloodlight.org/floodlight/}
\BIBentrySTDinterwordspacing

\bibitem{HassasYeganeh:2012:KFE:2342441.2342446}
S.~Hassas~Yeganeh and Y.~Ganjali, ``Kandoo: A framework for efficient and
  scalable offloading of control applications,'' in \emph{Proc. of Workshop on
  Hot Topics in Software Defined Networks}, 2012, pp. 19--24.

\bibitem{KARAKUS2017279}
M.~Karakus and A.~Durresi, ``A survey: Control plane scalability issues and
  approaches in software-defined networking,'' \emph{Comp. Net.}, vol. 112, pp.
  279 -- 293, 2017.

\bibitem{OKTIAN2017100}
Y.~E. Oktian, S.~Lee, H.~Lee, and J.~Lam, ``Distributed sdn controller system:
  A survey on design choice,'' \emph{Comp. Net.}, vol. 121, pp. 100 -- 111,
  2017.

\bibitem{McKeown:2008:OEI:1355734.1355746}
N.~McKeown, T.~Anderson, H.~Balakrishnan, G.~Parulkar, L.~Peterson, J.~Rexford,
  S.~Shenker, and J.~Turner, ``Openflow: Enabling innovation in campus
  networks,'' \emph{ACM Comput. Commun. Rev.}, vol.~38, no.~2, pp. 69--74,
  March 2008.

\bibitem{doria2010forwarding}
A.~Doria, J.~H. Salim, R.~Haas, H.~Khosravi, W.~Wang, L.~Dong, R.~Gopal, and
  J.~Halpern, ``Forwarding and control element separation (forces) protocol
  specification,'' {RFC} 5810, March 2010.

\bibitem{Song:2013:PFU:2491185.2491190}
H.~Song, ``Protocol-oblivious forwarding: Unleash the power of sdn through a
  future-proof forwarding plane,'' in \emph{Proc. of ACM Workshop on Hot Topics
  in Software Defined Networking}.\hskip 1em plus 0.5em minus 0.4em\relax ACM,
  2013, pp. 127--132.

\bibitem{pfaff2013open}
{Ben Pfaff and Bruce Davie}, ``{The open vSwitch database management
  protocol},'' {}, {RFC} 7047, December 2013.

\bibitem{parniewicz2014design}
D.~Parniewicz, R.~Doriguzzi~Corin, L.~Ogrodowczyk, M.~Rashidi~Fard, J.~Matias,
  M.~Gerola, V.~Fuentes, U.~Toseef, A.~Zaalouk, B.~Belter \emph{et~al.},
  ``Design and implementation of an openflow hardware abstraction layer,'' in
  \emph{Proc. of ACM Workshop on Distributed Cloud Computing}.\hskip 1em plus
  0.5em minus 0.4em\relax ACM, 2014, pp. 71--76.

\bibitem{inferringOFrules_Iqbal}
P.~C. Lin, P.~C. Li, and V.~L. Nguyen, ``Inferring openflow rules by active
  probing in software-defined networks,'' in \emph{Proc. of Int. Conf. on
  Advanced Communication Technology}, February 2017, pp. 415--420.

\bibitem{Jain:2013:BEG:2486001.2486019}
S.~Jain, A.~Kumar, S.~Mandal, J.~Ong \emph{et~al.}, ``B4: Experience with a
  globally-deployed software defined wan,'' in \emph{Proc. of ACM
  SIGCOMM}.\hskip 1em plus 0.5em minus 0.4em\relax ACM, 2013, pp. 3--14.

\bibitem{6305261}
D.~Kotani, K.~Suzuki, and H.~Shimonishi, ``A design and implementation of
  openflow controller handling ip multicast with fast tree switching,'' in
  \emph{IEEE/IPSJ Int. Symposium on Applications and the Internet}, July 2012,
  pp. 60--67.

\bibitem{Nakao}
\BIBentryALTinterwordspacing
A.~Nakao, ``Flare: Open deeply programmable network node architecture,''
  [Accessed 15-January-2019]. [Online]. Available:
  \url{http://netseminar.stanford.edu/10_18_12.html}
\BIBentrySTDinterwordspacing

\bibitem{Reitblatt:2012:ANU:2342356.2342427}
M.~Reitblatt, N.~Foster, J.~Rexford, C.~Schlesinger, and D.~Walker,
  ``Abstractions for network update,'' in \emph{Proc. of ACM Conf. on
  Applications, Technologies, Architectures, and Protocols for Comp. Commun.},
  2012, pp. 323--334.

\bibitem{6126682}
D.~M.~F. Mattos, N.~C. Fernandes, V.~T. da~Costa, L.~P. Cardoso \emph{et~al.},
  ``Omni: Openflow management infrastructure,'' in \emph{Proc. of Int. Conf. on
  the Network of the Future}, November 2011, pp. 52--56.

\bibitem{Wang:2011:OSL:1972422.1972438}
R.~Wang, D.~Butnariu, and J.~Rexford, ``Openflow-based server load balancing
  gone wild,'' in \emph{Proc. of USENIX Conf. on Hot Topics in Management of
  Internet, Cloud, and Enterprise Networks and Services}.\hskip 1em plus 0.5em
  minus 0.4em\relax ACM, 2011.

\bibitem{Gember:2012:TSM:2390231.2390233}
A.~Gember, P.~Prabhu, Z.~Ghadiyali, and A.~Akella, ``Toward software-defined
  middlebox networking,'' in \emph{Proc. of Workshop on Hot Topics in
  Networks}.\hskip 1em plus 0.5em minus 0.4em\relax ACM, 2012, pp. 7--12.

\bibitem{Gibb2012InitialTO}
\BIBentryALTinterwordspacing
G.~Gibb, H.~Zeng, and N.~McKeown, ``Initial thoughts on custom network
  processing via waypoint services,'' 2012, [Accessed 15-January, 2019].
  [Online]. Available: \url{https://www.semanticscholar.org/}
\BIBentrySTDinterwordspacing

\bibitem{ETSI}
\BIBentryALTinterwordspacing
ETSI, ``Network functions virtualisation (nfv),'' 2013, [Accessed 15-January,
  2019]. [Online]. Available:
  \url{https://portal.etsi.org/nfv/nfv_white_paper2.pdf}
\BIBentrySTDinterwordspacing

\bibitem{6702549}
W.~John, K.~Pentikousis, G.~Agapiou, E.~Jacob, M.~Kind, A.~Manzalini, F.~Risso,
  D.~Staessens, R.~Steinert, and C.~Meirosu, ``Research directions in network
  service chaining,'' in \emph{Proc. of IEEE SDN for Future Networks and
  Services}, November 2013, pp. 1--7.

\bibitem{Nayak:2009:RDA:1592681.1592684}
A.~K. Nayak, A.~Reimers, N.~Feamster, and R.~Clark, ``Resonance: Dynamic access
  control for enterprise networks,'' in \emph{Proc. of Workshop on Research on
  Enterprise Networking}.\hskip 1em plus 0.5em minus 0.4em\relax ACM, 2009, pp.
  11--18.

\bibitem{Khurshid:2012:VVN:2342441.2342452}
A.~Khurshid, W.~Zhou, M.~Caesar, and P.~B. Godfrey, ``Veriflow: Verifying
  network-wide invariants in real time,'' in \emph{Proc. of Workshop on Hot
  Topics in Software Defined Networks}.\hskip 1em plus 0.5em minus 0.4em\relax
  ACM, 2012, pp. 49--54.

\bibitem{6089085}
G.~Yao, J.~Bi, and P.~Xiao, ``Source address validation solution with
  openflow/nox architecture,'' in \emph{Proc. of IEEE Int. Conf. on Network
  Protocols}, October 2011, pp. 7--12.

\bibitem{Jafarian:2012:ORH:2342441.2342467}
J.~H. Jafarian, E.~Al-Shaer, and Q.~Duan, ``Openflow random host mutation:
  Transparent moving target defense using software defined networking,'' in
  \emph{Proc. of Workshop on Hot Topics in Software Defined Networks}.\hskip
  1em plus 0.5em minus 0.4em\relax ACM, 2012, pp. 127--132.

\bibitem{5689156}
C.~YuHunag, T.~MinChi, C.~YaoTing, C.~YuChieh, and C.~YanRen, ``A novel design
  for future on-demand service and security,'' in \emph{Proc. of Int. IEEE
  Conf. on Comm. Technology}, November 2010, pp. 385--388.

\bibitem{5735752}
R.~Braga, E.~Mota, and A.~Passito, ``Lightweight ddos flooding attack detection
  using nox/openflow,'' in \emph{Proc. of IEEE Local Computer Network Conf.},
  October 2010, pp. 408--415.

\bibitem{Porras:2012:SEK:2342441.2342466}
P.~Porras, S.~Shin, V.~Yegneswaran, M.~Fong, M.~Tyson, and G.~Gu, ``A security
  enforcement kernel for openflow networks,'' in \emph{Proc. of Workshop on Hot
  Topics in Software Defined Networks}.\hskip 1em plus 0.5em minus 0.4em\relax
  ACM, 2012, pp. 121--126.

\bibitem{Sharafat:2011:MMV:2018436.2018516}
A.~R. Sharafat, S.~Das, G.~Parulkar, and N.~McKeown, ``Mpls-te and mpls vpns
  with openflow,'' in \emph{Proc. of ACM SIGCOMM}.\hskip 1em plus 0.5em minus
  0.4em\relax ACM, 2011, pp. 452--453.

\bibitem{6066002}
S.~Azodolmolky, R.~Nejabati, E.~Escalona, R.~Jayakumar, N.~Efstathiou, and
  D.~Simeonidou, ``Integrated openflow x2014; gmpls control plane: An overlay
  model for software defined packet over optical networks,'' in \emph{Proc. of
  European Conf. and Exhibition on Optical Comm.}, September 2011, pp. 1--3.

\bibitem{Gutz:2012:SIS:2342441.2342458}
S.~Gutz, A.~Story, C.~Schlesinger, and N.~Foster, ``Splendid isolation: A slice
  abstraction for software-defined networks,'' in \emph{Proc. Workshop on Hot
  Topics in Software Defined Networks}.\hskip 1em plus 0.5em minus 0.4em\relax
  ACM, 2012, pp. 79--84.

\bibitem{Ferguson:2012:HPS:2342441.2342450}
A.~D. Ferguson, A.~Guha, C.~Liang, R.~Fonseca, and S.~Krishnamurthi,
  ``Hierarchical policies for software defined networks,'' in \emph{Proc.
  Workshop on Hot Topics in Software Defined Networks}, ser. HotSDN '12.\hskip
  1em plus 0.5em minus 0.4em\relax ACM, 2012, pp. 37--42.

\bibitem{6461196}
M.~Banikazemi, D.~Olshefski, A.~Shaikh, J.~Tracey, and G.~Wang, ``Meridian: an
  sdn platform for cloud network services,'' \emph{IEEE Commun. Mag.}, vol.~51,
  no.~2, pp. 120--127, February 2013.

\bibitem{openstackfoundation}
\BIBentryALTinterwordspacing
``The openstack foundatation.'' 2013, [Accessed 15-January, 2019]. [Online].
  Available: \url{http://www.openstack.org/}
\BIBentrySTDinterwordspacing

\bibitem{Nascimento:2010:QPQ:1851182.1851252}
M.~R. Nascimento, C.~E. Rothenberg, M.~R. Salvador, and M.~F. Magalh\~{a}es,
  ``Quagflow: Partnering quagga with openflow,'' in \emph{Proc. of ACM
  SIGCOMM}.\hskip 1em plus 0.5em minus 0.4em\relax ACM, 2010, pp. 441--442.

\bibitem{Nascimento:2011:VRS:2002396.2002405}
M.~R. Nascimento, C.~E. Rothenberg, M.~R. Salvador, C.~N.~A. Corr\^{e}a, S.~C.
  de~Lucena, and M.~F. Magalh\~{a}es, ``Virtual routers as a service: The
  routeflow approach leveraging software-defined networks,'' in \emph{Proc. of
  Int. Conf. on Future Internet Technologies}.\hskip 1em plus 0.5em minus
  0.4em\relax ACM, 2011, pp. 34--37.

\bibitem{6211892}
R.~Bennesby, P.~Fonseca, E.~Mota, and A.~Passito, ``An inter-as routing
  component for software-defined networks,'' in \emph{Proc. of IEEE Network
  Operations and Management Symposium}, April 2012, pp. 138--145.

\bibitem{Caesar:2005:DIR:1251203.1251205}
M.~Caesar, D.~Caldwell, N.~Feamster, J.~Rexford, A.~Shaikh, and J.~van~der
  Merwe, ``Design and implementation of a routing control platform,'' in
  \emph{Proc. of Int. Conf. on Symposium on Networked Systems Design \&
  Implementation - Volume 2}, 2005, pp. 15--28.

\bibitem{Rothenberg:2012:RRC:2342441.2342445}
C.~E. Rothenberg, M.~R. Nascimento, M.~R. Salvador, C.~N.~A. Corr\^{e}a,
  S.~Cunha~de Lucena, and R.~Raszuk, ``Revisiting routing control platforms
  with the eyes and muscles of software-defined networking,'' in \emph{Proc. of
  Workshop on Hot Topics in Software Defined Networks}, ser. HotSDN '12.\hskip
  1em plus 0.5em minus 0.4em\relax ACM, 2012, pp. 13--18.

\bibitem{restapi}
\BIBentryALTinterwordspacing
``{REST API} for {N}etwork {E}ngineers.'' 2016, [Accessed 15-January-2019].
  [Online]. Available:
  \url{http://networkop.co.uk/blog/2016/01/01/rest-for-neteng/}
\BIBentrySTDinterwordspacing

\bibitem{Ferguson:2013:PNA:2534169.2486003}
A.~D. Ferguson, A.~Guha, C.~Liang, R.~Fonseca, and S.~Krishnamurthi,
  ``Participatory networking: An api for application control of sdns,''
  \emph{ACM Comput. Commun. Rev.}, vol.~43, no.~4, pp. 327--338, 2013.

\bibitem{CISCO}
\BIBentryALTinterwordspacing
W.~Stallings, ``Software-defined networks and openflow,'' \emph{The Internet
  Protocol Journal}, vol.~16, no.~1, 2018, [Accessed 15-January, 2019].
  [Online]. Available:
  \url{https://www.cisco.com/c/en/us/about/press/internet-protocol-journal/back-issues/table-contents-59/161-sdn.html}
\BIBentrySTDinterwordspacing

\bibitem{7890085}
C.~Park and D.~Shin, ``{VNF} management method using vnf group table in network
  function virtualization,'' in \emph{Proc. of Int. Conf. on Advanced
  Communication Technology}, February 2017, pp. 210--212.

\bibitem{8404837}
Y.~Demıral and M.~Demırcı, ``An investigation of hypervisor effect on
  virtual networks performance,'' in \emph{Proc. of Signal Processing and Comm.
  Applications Conference (SIU)}, May 2018, pp. 1--4.

\bibitem{doi:sensorvirtualization}
J.~Ko, B.-B. Lee, K.~Lee, S.~G. Hong, N.~Kim, and J.~Paek, ``Sensor
  virtualization module: Virtualizing iot devices on mobile smartphones for
  effective sensor data management,'' \emph{Int. Journal of Distributed Sensor
  Networks}, vol.~11, no.~10, p. 730762, 2015.

\bibitem{zeroconf}
S.~Chesire and D.~Steinberg, ``{Z}ero {C}onfiguration {N}etwork-{T}he
  {D}efinitive {G}uide,'' 2006, [Accessed 30-January-2019].

\bibitem{5416827}
P.~Evensen and H.~Meling, ``Sensewrap: A service oriented middleware with
  sensor virtualization and self-configuration,'' in \emph{Proc. of Int. Conf.
  on Intelligent Sensors, Sensor Networks and Information Processing (ISSNIP)},
  December 2009, pp. 261--266.

\bibitem{6722995}
Y.~J. Fan, Y.~H. Yin, L.~D. Xu, Y.~Zeng, and F.~Wu, ``Iot-based smart
  rehabilitation system,'' \emph{IEEE Transactions on Industrial Informatics},
  vol.~10, no.~2, pp. 1568--1577, 2014.

\bibitem{IoT1}
\BIBentryALTinterwordspacing
J.~Bradley, ``The internet of everything: Creating better experiences in
  unimaginable ways.'' 2013, [Accessed 16-January-2019]. [Online]. Available:
  \url{https://blogs.cisco.com/digital#more-131793}
\BIBentrySTDinterwordspacing

\bibitem{sheng2013survey}
Z.~Sheng, S.~Yang, Y.~Yu, A.~V. Vasilakos, J.~Mccann, and K.~Leung, ``{A survey
  on the IETF protocol suite for the internet of things: Standards, challenges,
  and opportunities},'' \emph{IEEE Wireless Commun.}, vol.~20, no.~6, pp.
  91--98, 2013.

\bibitem{withanage2014comparison}
C.~Withanage, R.~Ashok, C.~Yuen, and K.~Otto, ``{A comparison of the popular
  home automation technologies},'' in \emph{Innovative Smart Grid
  Technologies-Asia (ISGT Asia)}.\hskip 1em plus 0.5em minus 0.4em\relax IEEE,
  2014, pp. 600--605.

\bibitem{kinney2017ieee}
T.~Kivinen and P.~Kinney, ``{IEEE 802.15.4 Information Element for the IETF},''
  RFC 8137, May 2017.

\bibitem{zhou2009wireless}
Y.~Zhou, X.~Yang, L.~Wang, and Y.~Ying, ``A wireless design of low-cost
  irrigation system using zigbee technology,'' in \emph{Proc. of Int. Conf. on
  Networks Security, Wireless Comm. and Trusted Comp.}, vol.~1, April 2009, pp.
  572--575.

\bibitem{deng2015ieee}
C.~Shao, D.~Hui, R.~Pazhyannur, F.~Bari, and R.~Zhang, ``{IEEE 802.11 Medium
  Access Control (MAC) Profile for Control and Provisioning of Wireless Access
  Points (CAPWAP)},'' RFC 7494, Apr. 2015.

\bibitem{nieminen2015rfc}
J.~Nieminen, T.~Savolainen, M.~Isomaki, B.~Patil, Z.~Shelby, and C.~Gomez,
  ``{IPv6 over BLUETOOTH(R) Low Energy},'' RFC 7668, Oct. 2015.

\bibitem{bisdikian2001overview}
C.~Bisdikian, ``An overview of the bluetooth wireless technology,'' \emph{IEEE
  Commun. Mag.}, vol.~39, no.~12, pp. 86--94, December 2001.

\bibitem{shelby20116lowpan}
Z.~Shelby and C.~Bormann, \emph{{{6LoWPAN: The wireless embedded
  Internet}}}.\hskip 1em plus 0.5em minus 0.4em\relax John Wiley \& Sons, 2011,
  vol.~43.

\bibitem{kushalnagar2007ipv6}
N.~Kushalnagar, G.~Montenegro, and C.~Schumacher, ``Ipv6 over low-power
  wireless personal area networks (6lowpans): overview, assumptions, problem
  statement, and goals,'' {}, {RFC} 4919, August 2007.

\bibitem{rappaport2013millimeter}
T.~S. Rappaport, S.~Sun, R.~Mayzus, H.~Zhao, Y.~Azar, K.~Wang, G.~N. Wong,
  J.~K. Schulz, M.~Samimi, and F.~Gutierrez, ``{{Millimeter wave mobile
  communications for 5G cellular: It will work!}}'' \emph{IEEE Access}, vol.~1,
  pp. 335--349, 2013.

\bibitem{niyato2017wireless}
D.~Niyato, D.~I. Kim, M.~Maso, and Z.~Han, ``Wireless powered communication
  networks: Research directions and technological approaches,'' \emph{IEEE
  Wireless Commun.}, vol.~PP, no.~99, pp. 2--11, 2017.

\bibitem{7060643}
I.~Khan, F.~Belqasmi, R.~Glitho, N.~Crespi, M.~Morrow, and P.~Polakos,
  ``Wireless sensor network virtualization: A survey,'' \emph{IEEE Commun.
  Surveys Tuts.}, vol.~18, no.~1, pp. 553--576, 2016.

\bibitem{8089336}
J.~Pan and J.~McElhannon, ``Future edge cloud and edge computing for internet
  of things applications,'' \emph{IEEE Internet Things J.}, vol.~5, no.~1, pp.
  439--449, 2018.

\bibitem{8141874}
G.~A. Akpakwu, B.~J. Silva, G.~P. Hancke, and A.~M. Abu-Mahfouz, ``A survey on
  5g networks for the internet of things: Communication technologies and
  challenges,'' \emph{IEEE Access}, vol.~6, pp. 3619--3647, 2018.

\bibitem{8066287}
J.~H. Cox, J.~Chung, S.~Donovan, J.~Ivey, R.~J. Clark, G.~Riley, and H.~L.
  Owen, ``Advancing software-defined networks: A survey,'' \emph{IEEE Access},
  vol.~5, pp. 25\,487--25\,526, 2017.

\bibitem{7582463}
A.~H. Ngu, M.~Gutierrez, V.~Metsis, S.~Nepal, and Q.~Z. Sheng, ``Iot
  middleware: A survey on issues and enabling technologies,'' \emph{IEEE
  Internet Things J.}, vol.~4, no.~1, pp. 1--20, 2017.

\bibitem{Tayyaba:2017:SDN:3102304.3102319}
S.~K. Tayyaba, M.~A. Shah, O.~A. Khan, and A.~W. Ahmed, ``Software defined
  network based internet of things: A road ahead,'' in \emph{Proc. of Int.
  Conf. on Future Networks and Distributed Systems}.\hskip 1em plus 0.5em minus
  0.4em\relax ACM, 2017, pp. 15:1--15:8.

\bibitem{huawei1}
\BIBentryALTinterwordspacing
M.~Yun, ``Huawei agile network: A solution for the three major problems facing
  traditional networking,'' 2015, [Accessed 15-January, 2019]. [Online].
  Available:
  \url{http://e.huawei.com/hk/publications/global/ict_insights/hw_314355/industry%20focus/HW_314358}
\BIBentrySTDinterwordspacing

\bibitem{huawei}
\BIBentryALTinterwordspacing
``{H}uawei {EC}-{I}o{T} {S}olution,'' 2015, [Accessed 15-January, 2019].
  [Online]. Available:
  \url{https://e.huawei.com/us/solutions/technical/sdn/agile-iot}
\BIBentrySTDinterwordspacing

\bibitem{IoT_SDN_OFenabled}
A.~Desai, K.~S. Nagegowda, and T.~Ninikrishna, ``A framework for integrating
  iot and sdn using proposed of-enabled management device,'' in \emph{Proc. of
  Int. Conf. on Circuit, Power and Computing Technologies}, March 2016, pp.
  1--4.

\bibitem{IoTEcoSysIqbal}
.~Ogrodowczyk, B.~Belter, and M.~LeClerc, ``Iot ecosystem over programmable sdn
  infrastructure for smart city applications,'' in \emph{European Workshop on
  Software-Defined Networks}, 2016, pp. 49--51.

\bibitem{salman2015architecture}
O.~Salman, I.~Elhajj, A.~Kayssi, and A.~Chehab, ``An architecture for the
  {I}nternet of {T}hings with decentralized data and centralized control,'' in
  \emph{Proc. of IEEE/ACS Int. Conf. of Computer Systems and Applications},
  November 2015, pp. 1--8.

\bibitem{SIMECA_Iqbal}
B.~Nguyen, N.~Choi, M.~Thottan, and J.~V. der Merwe, ``Simeca: Sdn-based iot
  mobile edge cloud architecture,'' in \emph{Proc. of IFIP/IEEE Symposium on
  Integrated Network and Service Management (IM)}, May 2017, pp. 503--509.

\bibitem{qin2014software}
Z.~Qin, G.~Denker, C.~Giannelli, P.~Bellavista, and N.~Venkatasubramanian, ``A
  software defined networking architecture for the internet-of-things,'' in
  \emph{Proc. of IEEE Network Operations and Management Symposium}, May 2014,
  pp. 1--9.

\bibitem{martinezempowering}
\BIBentryALTinterwordspacing
P.~Martinez-Julia and A.~F. Skarmeta, ``Empowering the internet of things with
  software defined networking,'' \emph{FP7 European research project on the
  future Internet of Things}, [Accessed 15-January, 2019]. [Online]. Available:
  \url{https://www.semanticscholar.org/}
\BIBentrySTDinterwordspacing

\bibitem{li2015general}
J.~Li, E.~Altman, and C.~Touati, ``A general {SDN}-based {IoT} framework with
  {NVF} implementation,'' \emph{ZTE communications}, vol.~13, no.~3, pp.
  42--45, 2015.

\bibitem{li2016sdn}
Y.~Li, X.~Su, J.~Riekki, T.~Kanter, and R.~Rahmani, ``A sdn-based architecture
  for horizontal internet of things services,'' in \emph{IEEE Commumn.
  Conf.}\hskip 1em plus 0.5em minus 0.4em\relax IEEE, 2016, pp. 1--7.

\bibitem{SDN_IoT_architecture_with_NFV}
M.~Ojo, D.~Adami, and S.~Giordano, ``A sdn-iot architecture with nfv
  implementation,'' in \emph{IEEE Globecom Workshops (GC Wkshps)}, December
  2016, pp. 1--6.

\bibitem{NoviFlow}
\BIBentryALTinterwordspacing
``{N}ovi{F}low,'' 2018, [Accessed 15-January, 2019]. [Online]. Available:
  \url{https://noviflow.com/}
\BIBentrySTDinterwordspacing

\bibitem{8370397}
S.~Asadollahi, B.~Goswami, and M.~Sameer, ``Ryu controller's scalability
  experiment on software defined networks,'' in \emph{Proc. of IEEE Int.
  Conf.on Current Trends in Advanced Computing}, February 2018, pp. 1--5.

\bibitem{libelium}
\BIBentryALTinterwordspacing
``Libelium,'' 2018, [Accessed 15-January, 2019]. [Online]. Available:
  \url{http://www.libelium.com/}
\BIBentrySTDinterwordspacing

\bibitem{SpirentSTC}
\BIBentryALTinterwordspacing
``Spirent,'' 2018, [Accessed 15-January, 2019]. [Online]. Available:
  \url{https://www.spirent.com/}
\BIBentrySTDinterwordspacing

\bibitem{WILLE2015629}
E.~C. Wille and J.~R. Hoffmann, ``Genius – a genetic scheduling algorithm for
  high-performance switches,'' \emph{AEU - International Journal of Electronics
  and Communications}, vol.~69, no.~3, pp. 629 -- 635, 2015.

\bibitem{QualNet}
\BIBentryALTinterwordspacing
``Scalable network technologies,'' 2018, [Accessed 15-January, 2019]. [Online].
  Available:
  \url{https://web.scalable-networks.com/qualnet-network-simulator-software}
\BIBentrySTDinterwordspacing

\bibitem{6838332}
Z.~Qin, L.~Iannario, C.~Giannelli, P.~Bellavista, G.~Denker, and
  N.~Venkatasubramanian, ``Mina: A reflective middleware for managing dynamic
  multinetwork environments,'' in \emph{Proc. of IEEE Network Operations and
  Management Symposium}, May 2014, pp. 1--4.

\bibitem{Seo:2008:GAS:1363686.1364121}
J.-H. Seo, Y.-H. Kim, H.-B. Ryou, and S.-J. Kang, ``A genetic algorithm for
  sensor deployment based on two-dimensional operators,'' in \emph{Proc. of
  Symposium on Applied Computing}.\hskip 1em plus 0.5em minus 0.4em\relax ACM,
  2008, pp. 1812--1813.

\bibitem{PhantomNet}
A.~Banerjee, J.~Cho, E.~Eide, J.~Duerig, B.~Nguyen, R.~Ricci, J.~Van~der Merwe,
  K.~Webb, and G.~Wong, ``Phantomnet: Research infrastructure for mobile
  networking, cloud computing and software-defined networking,''
  \emph{GetMobile: Mobile Comp. and Comm.}, vol.~19, no.~2, pp. 28--33, 2015.

\bibitem{xu2017defending}
T.~Xu, D.~Gao, P.~Dong, H.~Zhang, C.~H. Foh, and H.~C. Chao, ``Defending
  against new-flow attack in sdn-based internet of things,'' \emph{IEEE
  Access}, vol.~5, pp. 3431--3443, 2017.

\bibitem{sandor2015resilience}
H.~Sándor, B.~Genge, and G.~Sebestyén-Pál, ``Resilience in the internet of
  things: The software defined networking approach,'' in \emph{Proc. of IEEE
  Int. Conf. on Intelligent Computer Communication and Processing}, September
  2015, pp. 545--552.

\bibitem{hesham2017simplified}
A.~Hesham, F.~Sardis, S.~Wong, T.~Mahmoodi, and M.~Tatipamula, ``A simplified
  network access control design and implementation for m2m communication using
  sdn,'' in \emph{Proc. of IEEE Wireless Comm. and Networking Conf. Workshops},
  March 2017, pp. 1--5.

\bibitem{bull2016flow}
P.~Bull, R.~Austin, E.~Popov, M.~Sharma, and R.~Watson, ``Flow based security
  for iot devices using an sdn gateway,'' in \emph{Proc. of IEEE Int. Conf. on
  Future Internet of Things and Cloud}, August 2016, pp. 157--163.

\bibitem{sivanathan2016low}
A.~Sivanathan, D.~Sherratt, H.~H. Gharakheili, V.~Sivaraman, and A.~Vishwanath,
  ``Low-cost flow-based security solutions for smart-home iot devices,'' in
  \emph{Proc. of IEEE Int. Conf. on Advanced Networks and Telecommunications
  Systems}, November 2016, pp. 1--6.

\bibitem{sivaraman2015network}
V.~Sivaraman, H.~H. Gharakheili, A.~Vishwanath, R.~Boreli, and O.~Mehani,
  ``Network-level security and privacy control for smart-home iot devices,'' in
  \emph{Proc. of IEEE Int. Conf. on Wireless and Mobile Computing, Networking
  and Communications}, October 2015, pp. 163--167.

\bibitem{HostBasedIntrusion_Iqbal}
M.~Nobakht, V.~Sivaraman, and R.~Boreli, ``A host-based intrusion detection and
  mitigation framework for smart home iot using openflow,'' in \emph{Proc. of
  Int. Conf. on Availability, Reliability and Security}, August 2016, pp.
  147--156.

\bibitem{flauzac2015sdn}
O.~Flauzac, C.~González, A.~Hachani, and F.~Nolot, ``Sdn based architecture
  for iot and improvement of the security,'' in \emph{Proc. of IEEE Int. Conf.
  on Advanced Information Networking and Applications Workshops}, March 2015,
  pp. 688--693.

\bibitem{gonzalez2016sdn}
C.~Gonzalez, S.~M. Charfadine, O.~Flauzac, and F.~Nolot, ``Sdn-based security
  framework for the iot in distributed grid,'' in \emph{Proc. of Int.
  Multidisciplinary Conf. on Computer and Energy Science (SpliTech)}, July
  2016, pp. 1--5.

\bibitem{7785005}
F.~A. Shuhaimi, M.~Jose, and A.~V. Singh, ``Software defined network as
  solution to overcome security challenges in iot,'' in \emph{Proc. of Int.l
  Conf. on Reliability, Infocom Technologies and Optimization (Trends and
  Future Directions)}, September 2016, pp. 491--496.

\bibitem{li2017securing}
C.~Li, Z.~Qin, E.~Novak, and Q.~Li, ``Securing sdn infrastructure of iot-fog
  network from mitm attacks,'' \emph{IEEE Internet Things J.}, vol.~PP, no.~99,
  2017.

\bibitem{chakrabarty2015black}
S.~Chakrabarty, D.~W. Engels, and S.~Thathapudi, ``Black sdn for the internet
  of things,'' in \emph{Proc. of IEEE Int. Conf. on Mobile Ad Hoc and Sensor
  Systems}, October 2015, pp. 190--198.

\bibitem{802.1x_rfc}
\BIBentryALTinterwordspacing
``802.1x: Port-based network access control.'' 2010, [Accessed 15-January,
  2019]. [Online]. Available: \url{https://1.ieee802.org/security/802-1x/}
\BIBentrySTDinterwordspacing

\bibitem{bull2015pre}
P.~Bull, R.~Austin, and M.~Sharma, ``Pre-emptive flow installation for internet
  of things devices within software defined networks,'' in \emph{Proc. of Int.
  Conf. on Future Internet of Things and Cloud}, August 2015, pp. 124--130.

\bibitem{OpFlexProtocol}
\BIBentryALTinterwordspacing
``{OpFlex: An Open Source Approach},'' [Accessed 19-January-2019]. [Online].
  Available:
  \url{https://www.cisco.com/c/en/us/solutions/collateral/data-center-virtualization/application-centric-infrastructure/white-paper-c11-731304.html}
\BIBentrySTDinterwordspacing

\bibitem{maninthemiddle}
P.~Mutton, ``{95 percentage of HTTPS servers vulnerable to trivial MITM
  attacks},'' 2016.

\bibitem{lantz2010network}
B.~Lantz, B.~Heller, and N.~McKeown, ``A network in a laptop: Rapid prototyping
  for software-defined networks,'' in \emph{Proc. of ACM Workshop on Hot Topics
  in Networks}, ser. Hotnets-IX.\hskip 1em plus 0.5em minus 0.4em\relax ACM,
  2010, pp. 19:1--19:6.

\bibitem{hakiri2015publish}
A.~Hakiri, P.~Berthou, A.~Gokhale, and S.~Abdellatif,
  ``Publish/subscribe-enabled software defined networking for efficient and
  scalable iot communications,'' \emph{IEEE Commun. Mag.}, vol.~53, no.~9, pp.
  48--54, September 2015.

\bibitem{bera2016soft}
S.~Bera, S.~Misra, S.~K. Roy, and M.~S. Obaidat, ``Soft-wsn: Software-defined
  wsn management system for iot applications,'' \emph{IEEE Systems Journal},
  vol.~PP, no.~99, pp. 1--8, 2016.

\bibitem{yiakoumis2011slicing}
Y.~Yiakoumis, K.-K. Yap, S.~Katti, G.~Parulkar, and N.~McKeown, ``Slicing home
  networks,'' in \emph{Proc. of ACM Workshop on Home Networks}.\hskip 1em plus
  0.5em minus 0.4em\relax ACM, 2011, pp. 1--6.

\bibitem{7543778}
M.~Tortonesi, J.~Michaelis, A.~Morelli, N.~Suri, and M.~A. Baker, ``Spf: An
  sdn-based middleware solution to mitigate the iot information explosion,'' in
  \emph{Proc. of IEEE Symposium on Computers and Communication}, June 2016, pp.
  435--442.

\bibitem{fichera2017experimenting}
S.~Fichera, M.~Gharbaoui, P.~Castoldi, B.~Martini, and A.~Manzalini, ``On
  experimenting 5g: Testbed set-up for sdn orchestration across network cloud
  and iot domains,'' in \emph{Proc. of IEEE Conf. on Network Softwarization},
  July 2017, pp. 1--6.

\bibitem{sherwood2009flowvisor}
R.~Sherwood, G.~Gibb, K.-K. Yap, G.~Appenzeller, M.~Casado, N.~McKeown, and
  G.~Parulkar, ``Flowvisor: A network virtualization layer,'' \emph{OpenFlow
  Switch Consortium, Tech. Rep}, vol.~1, p. 132, 2009.

\bibitem{batalle2013implementation}
J.~Batalle, J.~F. Riera, E.~Escalona, and J.~A. Garcia-Espin, ``On the
  implementation of nfv over an openflow infrastructure: Routing function
  virtualization,'' in \emph{Future Networks and Services (SDN4FNS), IEEE SDN
  for}.\hskip 1em plus 0.5em minus 0.4em\relax IEEE, 2013, pp. 1--6.

\bibitem{du2016context}
P.~Du, P.~Putra, S.~Yamamoto, and A.~Nakao, ``A context-aware iot architecture
  through software-defined data plane,'' in \emph{Proc. of IEEE Region 10
  Symposium}, May 2016, pp. 315--320.

\bibitem{Edison}
\BIBentryALTinterwordspacing
Intel, ``Create prototypes and get to market faster using intel edison
  technology,'' [Accessed 15-January, 2019]. [Online]. Available:
  \url{https://software.intel.com/en-us/iot/hardware/edison/documentation}
\BIBentrySTDinterwordspacing

\bibitem{balon2012mobile}
M.~Balon and B.~Liau, ``Mobile virtual network operator,'' in \emph{Proc. of
  Int. Telecommunications Network Strategy and Planning Symposium}, October
  2012, pp. 1--6.

\bibitem{vilalta2016end}
R.~Vilalta, A.~Mayoral, D.~Pubill, R.~Casellas, R.~Martínez, J.~Serra,
  C.~Verikoukis, and R.~Muñoz, ``End-to-end sdn orchestration of iot services
  using an sdn/nfv-enabled edge node,'' in \emph{Proc. of Optical Fiber Comm.
  Conf. and Exhibition}, March 2016, pp. 1--3.

\bibitem{salman2015edge}
O.~Salman, I.~Elhajj, A.~Kayssi, and A.~Chehab, ``Edge computing enabling the
  internet of things,'' in \emph{Proc. of IEEE World Forum on Internet of
  Things}, December 2015, pp. 603--608.

\bibitem{maksymyuk2017iot}
T.~Maksymyuk, S.~Dumych, M.~Brych, D.~Satria, and M.~Jo, ``An iot based
  monitoring framework for software defined 5g mobile networks,'' in
  \emph{Proc. of Int. Conf. on Ubiquitous Information Management and
  Communication}.\hskip 1em plus 0.5em minus 0.4em\relax ACM, 2017, pp.
  105:1--105:4.

\bibitem{Zhang:2016:OPH:2940147.2940155}
W.~Zhang, G.~Liu, W.~Zhang, N.~Shah, P.~Lopreiato, G.~Todeschi,
  K.~Ramakrishnan, and T.~Wood, ``Opennetvm: A platform for high performance
  network service chains,'' in \emph{Proc. of Int. Conf. Workshop on Hot Topics
  in Middleboxes and Network Function Virtualization}.\hskip 1em plus 0.5em
  minus 0.4em\relax ACM, 2016, pp. 26--31.

\bibitem{8114476}
P.~Massonet, L.~Deru, A.~Achour, S.~Dupont, L.~M. Croisez, A.~Levin, and
  M.~Villari, ``Security in lightweight network function virtualisation for
  federated cloud and iot,'' in \emph{Proc. of Int. Conf. on Future Internet of
  Things and Cloud (FiCloud)}, August 2017, pp. 148--154.

\bibitem{8432333}
M.~Al-Shaboti, I.~Welch, A.~Chen, and M.~A. Mahmood, ``Towards secure smart
  home iot: Manufacturer and user network access control framework,'' in
  \emph{Proc. of Int. Conf. on Advanced Information Networking and
  Applications}, May 2018, pp. 892--899.

\bibitem{FLARE}
\BIBentryALTinterwordspacing
``Flare: Open deeply programmable network node architecture.'' 2018, [Accessed
  15-January, 2019]. [Online]. Available:
  \url{http://netseminar.stanford.edu/seminars/10_18_12.pdf}
\BIBentrySTDinterwordspacing

\bibitem{GEANT}
\BIBentryALTinterwordspacing
``Geant project: Software defined networking.'' 2017, [Accessed 15-January,
  2019]. [Online]. Available:
  \url{https://geant3plus.archive.geant.net/opencall/SDN/Pages/Home.aspx}
\BIBentrySTDinterwordspacing

\bibitem{DPDK}
\BIBentryALTinterwordspacing
``Data plane development kit.'' [Accessed 15-January, 2019]. [Online].
  Available: \url{http://www.dpdk.org/}
\BIBentrySTDinterwordspacing

\bibitem{8110222}
S.~Din, M.~M. Rathore, A.~Ahmad, A.~Paul, and M.~Khan, ``Sdiot: Software
  defined internet of thing to analyze big data in smart cities,'' in
  \emph{Proc. of Int. Conf. on Local Computer Networks Workshops (LCN
  Workshops)}, October 2017, pp. 175--182.

\bibitem{7986623}
S.~Guizani, ``Internet-of-things (iot) feasibility applications in information
  centric networking system,'' in \emph{Proc. of Int. Wireless Comm. and Mobile
  Comp. Conf. (IWCMC)}, June 2017, pp. 2192--2197.

\bibitem{8308426}
A.~Kalghoum, S.~M. Gammar, and L.~A. Saidane, ``Towards a novel forwarding
  strategy for named data networking based on sdn and bloom filter,'' in
  \emph{Proc. of IEEE/ACS Int. Conf. on Comp. Sys. and Applications (AICCSA)},
  October 2017, pp. 1198--1204.

\bibitem{liu2015software}
J.~Liu, Y.~Li, M.~Chen, W.~Dong, and D.~Jin, ``Software-defined internet of
  things for smart urban sensing,'' \emph{IEEE Commun. Mag.}, vol.~53, no.~9,
  pp. 55--63, September 2015.

\bibitem{xu2016toward}
K.~Xu, X.~Wang, W.~Wei, H.~Song, and B.~Mao, ``Toward software defined smart
  home,'' \emph{IEEE Commun. Mag.}, vol.~54, no.~5, pp. 116--122, 2016.

\bibitem{7324414}
P.~Hu, ``A system architecture for software-defined industrial internet of
  things,'' in \emph{Proc. of IEEE Int. Conf. on Ubiquitous Wireless
  Broadband}, October 2015, pp. 1--5.

\bibitem{7467436}
J.~Wan, S.~Tang, Z.~Shu, D.~Li, S.~Wang, M.~Imran, and A.~V. Vasilakos,
  ``Software-defined industrial internet of things in the context of industry
  4.0,'' \emph{IEEE Sensors Journal}, vol.~16, no.~20, pp. 7373--7380, 2016.

\bibitem{6984208}
S.~Nastic, S.~Sehic, D.~H. Le, H.~L. Truong, and S.~Dustdar, ``Provisioning
  software-defined iot cloud systems,'' in \emph{Proc. of Int.l Conf. on Future
  Internet of Things and Cloud}, August 2014, pp. 288--295.

\bibitem{Kathiravelu:2015:CMP:2836127.2836132}
P.~Kathiravelu, L.~Sharifi, and L.~Veiga, ``Cassowary: Middleware platform for
  context-aware smart buildings with software-defined sensor networks,'' in
  \emph{Workshop on Middleware for Context-Aware Applications in the IoT},
  2015, pp. 1--6.

\bibitem{wu2015ubiflow}
D.~Wu, D.~I. Arkhipov, E.~Asmare, Z.~Qin, and J.~A. McCann, ``Ubiflow: Mobility
  management in urban-scale software defined iot,'' in \emph{Proc. of IEEE
  Conf. on Comp.Comm.}, April 2015, pp. 208--216.

\bibitem{jararweh2015sdiot}
Y.~Jararweh, M.~Al-Ayyoub, A.~Darabseh, E.~Benkhelifa, M.~Vouk, and A.~Rindos,
  ``Sdiot: a software defined based internet of things framework,''
  \emph{Journal of Ambient Intelligence and Humanized Computing}, vol.~6,
  no.~4, pp. 453--461, 2015.

\bibitem{SDIoTSecurity_Iqbal}
O.~Salman, I.~Elhajj, A.~Chehab, and A.~Kayssi, ``Software defined iot security
  framework,'' in \emph{Proc. of Int. Conf. on Software Defined Systems}, May
  2017, pp. 75--80.

\bibitem{SDSecurityExpFramework_Iqbal}
A.~Darabseh, M.~Al-Ayyoub, Y.~Jararweh, E.~Benkhelifa, M.~Vouk, and A.~Rindos,
  ``Sdsecurity: A software defined security experimental framework,'' in
  \emph{Proc. of Int. Conf. on Comm. Workshop}, June 2015, pp. 1871--1876.

\bibitem{6224230}
S.~Dawson-Haggerty, J.~Ortiz, J.~Trager, D.~Culler, and R.~H. Katz, ``Energy
  savings and the “software-defined” building,'' \emph{IEEE Design Test of
  Computers}, vol.~29, no.~4, pp. 56--57, 2012.

\bibitem{catbird}
\BIBentryALTinterwordspacing
``Catbird security software suite.'' 2018, [Accessed 19-January-2019].
  [Online]. Available:
  \url{https://www.sdxcentral.com/products/catbird-security-software-suite/}
\BIBentrySTDinterwordspacing

\bibitem{yin2012}
\BIBentryALTinterwordspacing
H.~Yin, H.~Xie, T.~Tsou, D.~Lopez, P.~Aranda, and R.~Sidi, ``{SDNi: A Message
  Exchange Protocol for Software Defined Networks (SDNS) across Multiple
  Domains},'' Internet Draft, Internet Engineering Task Force, June 2012.
  [Online]. Available: \url{http://tools.ietf.org/id/draft-yin-sdn-sdni-00.txt}
\BIBentrySTDinterwordspacing

\end{thebibliography}
\end{document}